\documentclass[aps,twocolumn,prc,superscriptaddress,noshowpacs,nofootinbib,noshowkeys,floatfix]{revtex4}
\usepackage[dvips]{graphics,graphicx}
\usepackage[colorlinks=true,linktocpage=true,linkcolor=blue,citecolor=blue]{hyperref}
\usepackage[usenames,dvipsnames]{color}
\usepackage{amsmath, amssymb}
\usepackage{multirow}
\usepackage{longtable}
\usepackage{color}
\usepackage{xcolor}
\usepackage{cleveref}
\usepackage[normalem]{ulem}  

\renewcommand\sout{\bgroup \color{blue} \ULdepth=-.5ex \ULset}

\begin{document}

\title{{\Large Effect of various particlization scenarios on anisotropic flow and particle production using UrQMD hybrid model}}

\author{Sumit Kumar Kundu\footnote{Corresponding author.}}
\email{sumitvedkundu@gmail.com}
\affiliation{Department of Physics, School of Basic Sciences, Indian Institute of Technology Indore, Indore 453552 India}

\author{Yoshini Bailung}
\affiliation{Department of Physics, School of Basic Sciences, Indian Institute of Technology Indore, Indore 453552 India}
\author{Sudhir Pandurang Rode}
\affiliation{Veksler and Baldin Laboratory of High Energy Physics, Joint Institute for Nuclear Research, Dubna 141980, Moscow region, Russian Federation}

\author{Partha Pratim Bhaduri}
\affiliation{Variable Energy Cyclotron Centre, HBNI, 1/AF Bidhan Nagar, Kolkata 700 064, India}
\author{Ankhi Roy}
\email{ankhi@iiti.ac.in}
\affiliation{Department of Physics, School of Basic Sciences, Indian Institute of Technology Indore, Indore 453552 India}

\begin{abstract}

We discuss the effect of various particlization scenarios available in the hybrid ultrarelativistic quantum molecular dynamics (UrQMD) event generator on different observables in non-central ($b$ $=$ 5--9 $fm$) Au + Au collisions in the beam energy range 1A-158A GeV. Particlization models switch fluid dynamic description to the transport description using various hypersurface criteria. In addition to particlization models, various equations-of-state (EoS) provided by the UrQMD hybrid model were employed. The observables examined in this paper include the excitation function of anisotropic coefficients such as directed ($v_{1}$) and elliptic flow ($v_{2}$), particle ratios of the species, and the shape of net-proton rapidity spectra at mid-rapidity. The results obtained here can help predict and compare the measurements provided by future experiments at the Facility for Antiproton and Ion Research (FAIR) and the Nuclotron-based Ion Collider fAcility (NICA) once the data becomes available. We also study the most suitable combination of the particlization model and EoS, which best describes the experimental measurements.
\end{abstract}

\maketitle
\section*{I. Introduction}
Relativistic collisions of heavy ions allow inspection of the phase structure of the strongly interacting matter produced in the laboratory over a wide range of temperatures and densities. One of the many objectives of these collisions is to locate the quantum chromodynamics (QCD) critical point and the phase transition to the deconfined matter as per the various QCD predictions. The high temperature and vanishing baryon chemical potential ($\mu_{B}$) region of the QCD phase diagram is pretty well explored in the experiments operating at the Large Hadron Collider (LHC)~\cite{lhc1,lhc2,lhc3} at CERN, Geneva and Relativistic Heavy Ion Collider (RHIC)~\cite{rhic1,rhic2} at BNL, USA. On the contrary, the realization of the QCD matter produced at high $\mu_{B}$ is somewhat limited, and investigation of which is being carried out at currently available RHIC Beam Energy Scan (BES) program and will be performed in upcoming experiments at facilities such as the Facility for Anti-proton and Ion Research (FAIR)~\cite{Ablyazimov:2017guv,Sturm:2010yit} at GSI, Germany and Nuclotron-based Ion Collider fAcility (NICA)~\cite{Kekelidze:2016wkp} at JINR, Russia. 

Due to technological limitations, previous experiments at these energies could not address problems associated with the rare probes. Thus, in addition to this, as well as for the optimal utilization of the new upcoming facilities, it is essential to investigate the data available from previous experiments at similar beam energy regimes and provide predictions for future experimental measurements. For the latter, a plethora of phenomenological models and simulation packages have been utilized to provide possible prospects for upcoming experiments.

In recent times, hybrid models have been beneficial for describing the evolution of strongly interacting matter in heavy-ion collisions. In the hybrid models, the transport approach to explain the non-equilibrium dynamics is connected with hydrodynamical description to describe the expansion of the locally thermalized fireball. Such a combination of approaches can be advantageous in investigating various observables to extract QCD medium properties. Anisotropic flow is one such observable which describes the collective expansion of the medium and is quantified using Fourier expansion of the final state azimuthal distribution of the produced particles, 
\begin{align*}
v_{n} = \big<\cos[n(\phi - \Psi)]\big>
\end{align*} 
where the azimuthal angle and reaction plane angle are denoted by $\phi$ and $\Psi$ respectively, whereas directed flow ($v_{1}$) and elliptic flow ($v_{2}$) are the first and second anisotropic coefficients.

Back to the hybrid models, UrQMD is the most popular event generator model, publicly available in a hybrid format. After completion of fluid dynamic evolution, the later stage is described using the transport approach as the medium is expected to be away from the equilibrium. The conversion of fluid-based description to the particle-based description is known as particlization, which is a technical terminology. The Cooper-Frye procedure~\cite{Cooper:1974mv} is used to evaluate the particle distributions on the boundary where this conversion is performed. It is essential here to note that this switching of description is not freeze-out, as in principle, after freeze-out, there should not be any rescattering and, therefore, no need to have transport description~\cite{Huovinen:2012is}. Various variants of freeze-out hypersurfaces or particlization models are available in the UrQMD hybrid model. One can aim to extract some insights about the medium by observing the response of different observables to these variants. The choice of particlization model can further affect the evolution of the particles after hadronization as well as freeze-out and hence observables, such as anisotropic flow, particle ratios, rapidity spectra, and so on. Since the corresponding physical stage to particlization is not available in actual evolution of the QCD matter, its optimization is necessary to describe the experimental measurements.

To look for the possible signatures of thermally equilibrated QCD medium, many experiments~\cite{Alver:2006wh,Aamodt:2010pa} at various energies have investigated the elliptic flow coefficient $v_{2}$ of the hadrons. Even in low energy collisions at various beam energy ranges~\cite{Bhaduri:2010wi,Sarkar:2017fuy,Auvinen:2013sba}, $v_{2}$ has been studied by employing microscopic transport models~\cite{Bass:1998ca,Bleicher:1999xi,Lin:2004en,Chen:2004vha}. The transition from out-of-plane to in-plane flow in terms of change of sign has been observed at low energies~\cite{LeFevre:2016vpp,Pinkenburg:1999ya}. In one of our previous works~\cite{Rode:2019pey}, the dependence of anisotropic flow has been studied using a hybrid UrQMD model for $6A-25A$ GeV beam energies with hadron gas (HG) and chiral EoS. In another work~\cite{Kundu:2021afz}, various flow coefficients of anisotropic momentum distribution and particle production have been analyzed for HG, chiral, and bag model (BM) EOS in the beam energy range $\rm E_{\rm Lab}$ $=$ $1A -158A$ GeV.

The directed flow, $v_{1}$, measures the deflection of the generated particles in the reaction plane. $v_{1}$ is worth investigating in relativistic nuclear collisions because of its sensitivity to longitudinal dynamics and the possibility of being formed before $v_{2}$~\cite{Nara:2016phs,Nara:1999dz,Konchakovski:2014gda}. Due to the softening of the underlying EoS, the magnitude of $v_{1}$ is projected to vanish in the region of the phase transition, making it an interesting observable for research at RHIC-BES, FAIR, and NICA energies. Recently, the anisotropic flow coefficients of hadrons and deuterons were studied at RHIC in STAR experiment~\cite{STAR:2020dav,STAR:2020hya} in Au + Au collisions. Various experiments have been conducted in this direction during the last few decades. For example, at Alternating Gradient Synchrotron (AGS)~\cite{Liu:2000am,Chung:2000ny,Chung:2001je} energies and less than that, the slope of $v_{1}$ at midrapidity range, which represents signal strength, displays linearity. Similar is not expected at higher beam energies because, in previous studies~\cite{Appelshauser:1997dg,Adams:2004bi,Back:2005pc}, it has been found that slope at midrapidity is different from that at beam rapidity at energies higher than Super Proton Synchrotron (SPS) energy. The so-called structure "wiggle" appears to be sensitive to the underlying EoS, according to hydrodynamic model estimates~\cite{Snellings:1999bt,Csernai:1999nf,Brachmann:1999xt}. Higher-order harmonics research has gotten much interest in recent years and is expected to reveal more information on the produced fireball. The fourth-order harmonic coefficient, $v_{4}$, is known to be sensitive to intrinsic $v_{2}$~\cite{Borghini:2005kd,Gombeaud:2009ye,Luzum:2010ae}; thus, it is worth looking at it for a wider range of beam energies, which has also been done using the jet AA microscopic transport model JAM~\cite{Nara:1999dz,Nara:2018ijw}. It contains important information on the collision dynamics anticipated by hydrodynamical simulations.

Several event generator models to simulate the relativistic heavy-ion collisions are at the disposal of physicists in the market. However, almost none of them can describe the experimental measurements under one framework. Thus, to better understand the precise measurements from upcoming experiments, it is imperative to address their shortcomings and fix them as much as possible. As part of this attempt, one can employ various options available in a particular model and try to extract the physics messages by performing qualitative or quantitative comparisons with existing experimental measurements.

In the present article, we aim to investigate the effect of various particlization scenarios using multiple particlization models in the UrQMD model for the available variety of the nuclear equations-of-state. For this, we simulate noncentral ($5 < b < 9$ fm) Au - Au collisions using the UrQMD model for beam energies ranging from 1A--158A GeV for different particlization models and EoS. The chosen beam energy range spans existing GSI-SIS energy of the HADES experiment up to top SPS energy. The corresponding $\rm <N_{part}>$ values in the selected impact parameter region of $5 < b < 9$ fm cover approximately 10–40$\%$ centrality class~\cite{Adare:2015bua}. Similar to our previous work, we compute various anisotropic coefficients and particle ratios and study net-proton rapidity distribution in the beam energy regime mentioned above. Our objective is to find the best possible combination among the various permutations of UrQMD variants which describe the data best.

\begin{figure*}[h!]
\includegraphics[scale=0.295]{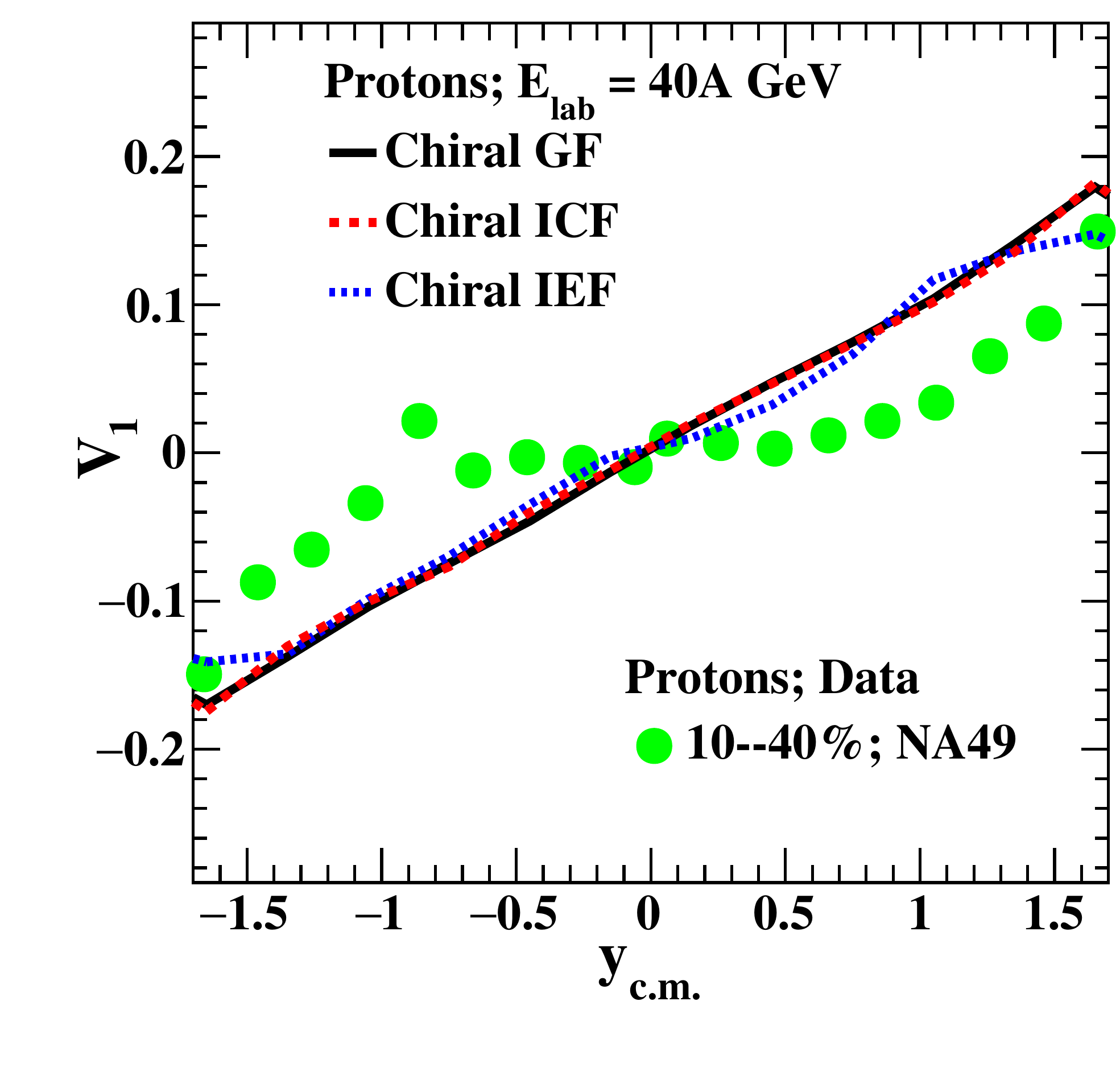}
\includegraphics[scale=0.295]{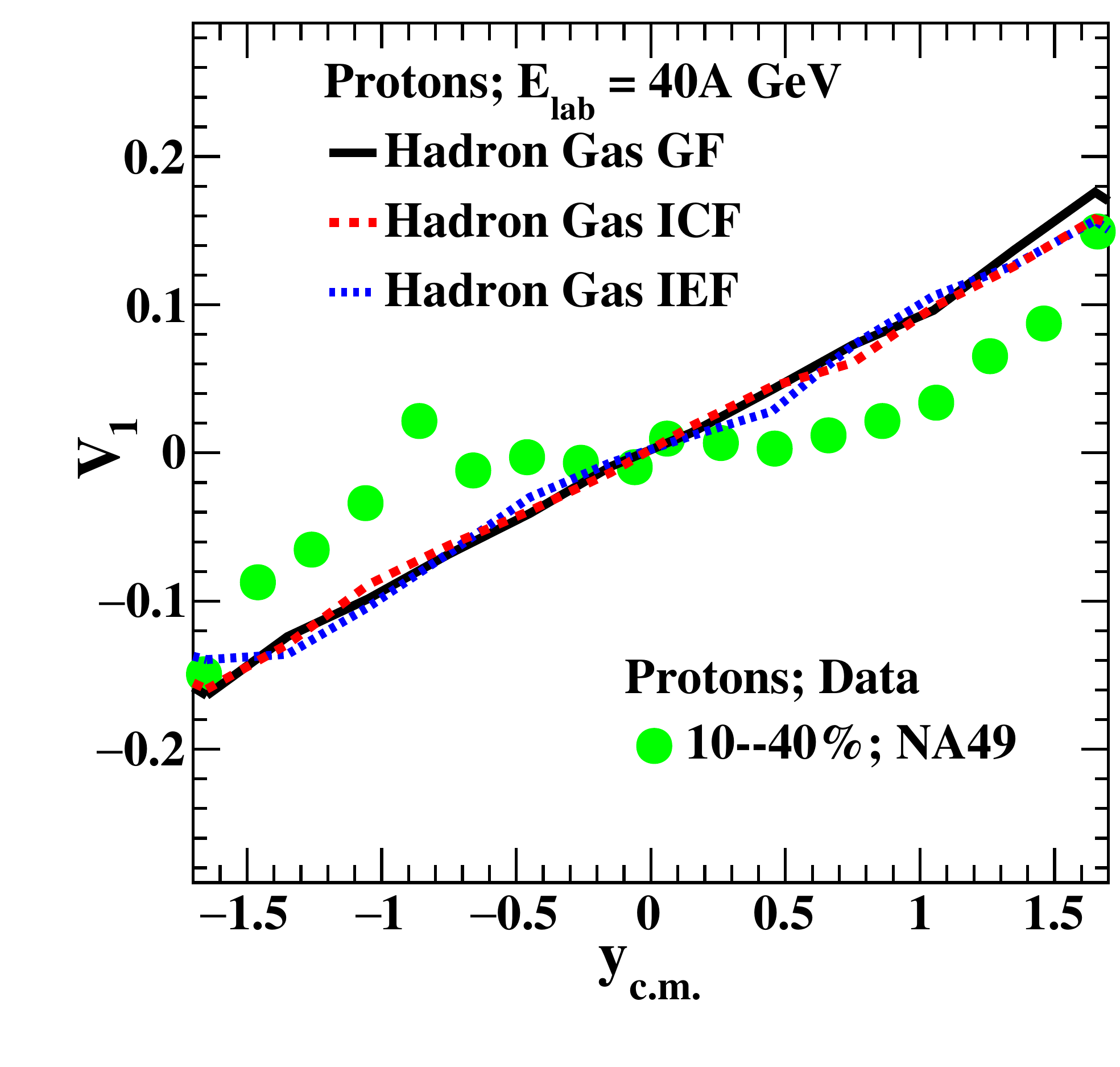}
\includegraphics[scale=0.295]{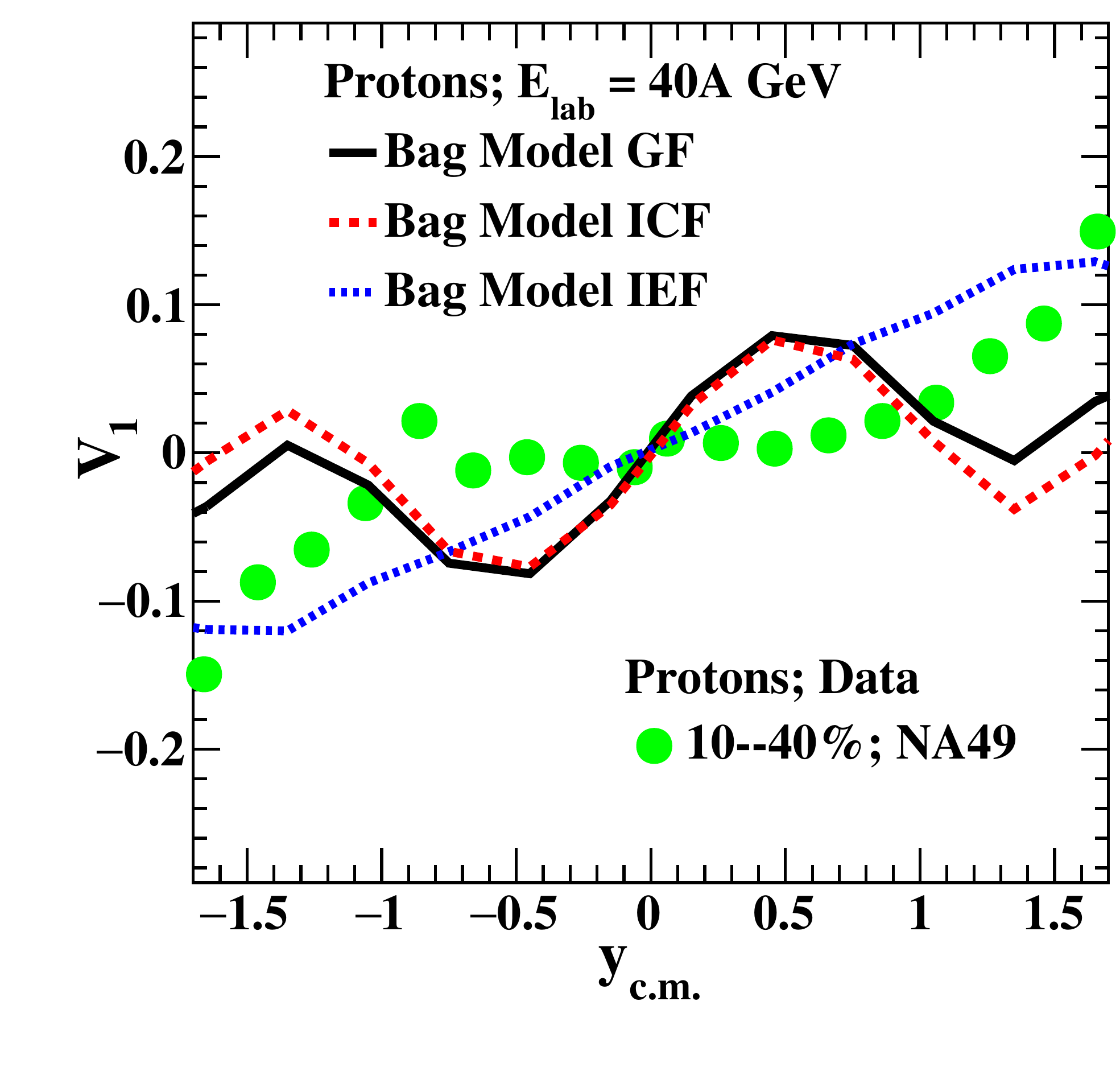}\\
\includegraphics[scale=0.295]{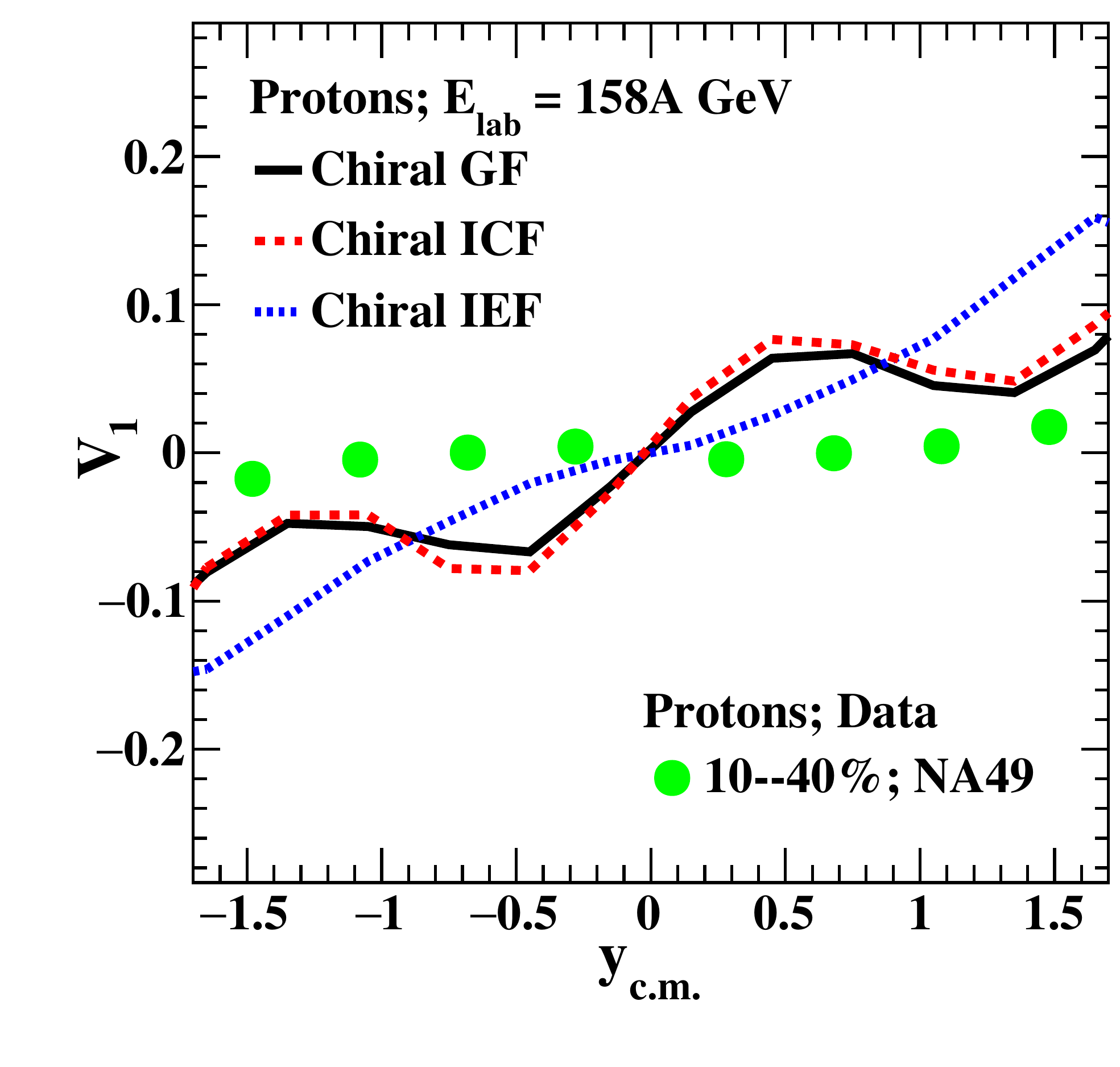}
\includegraphics[scale=0.295]{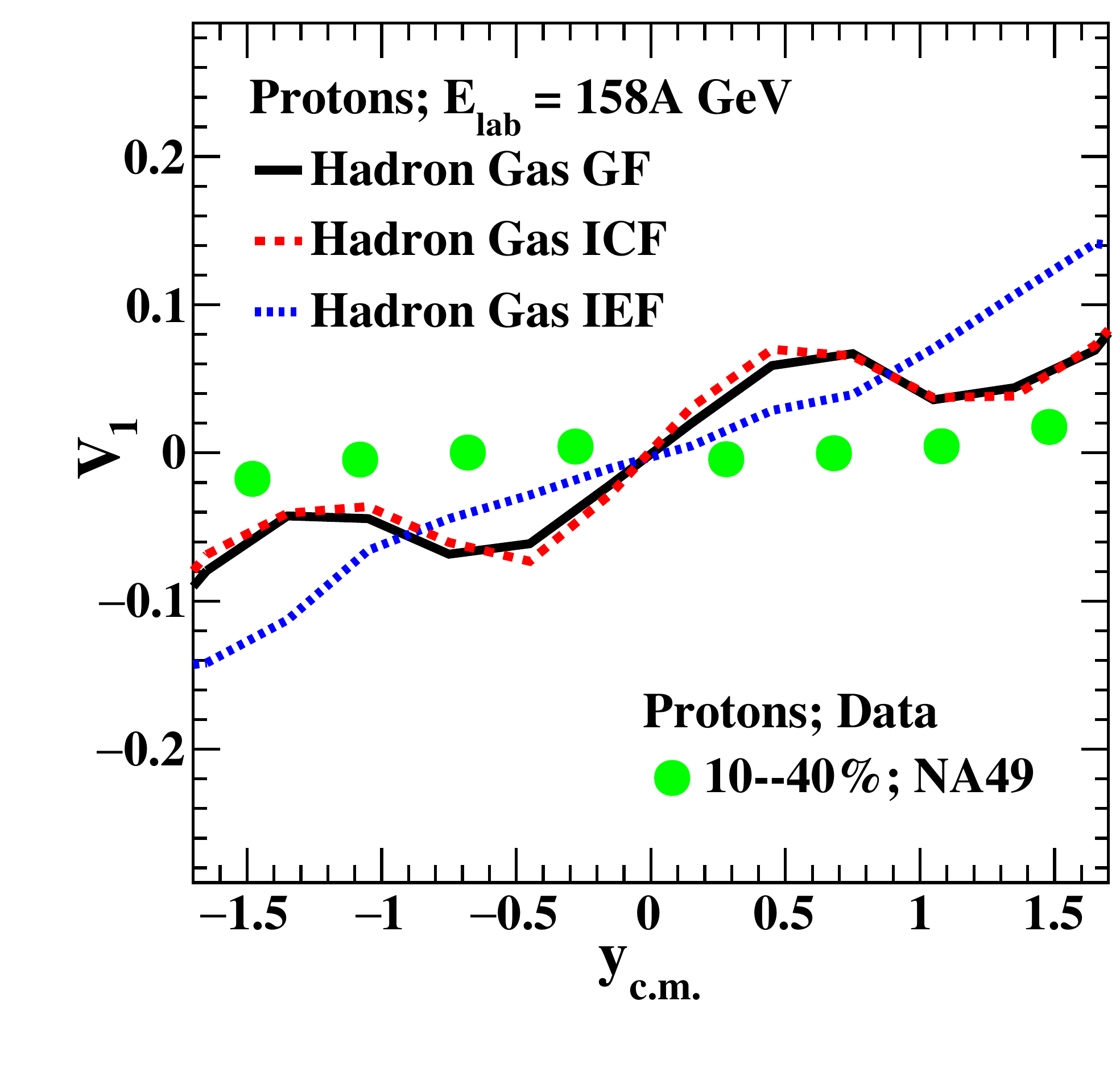}
\includegraphics[scale=0.295]{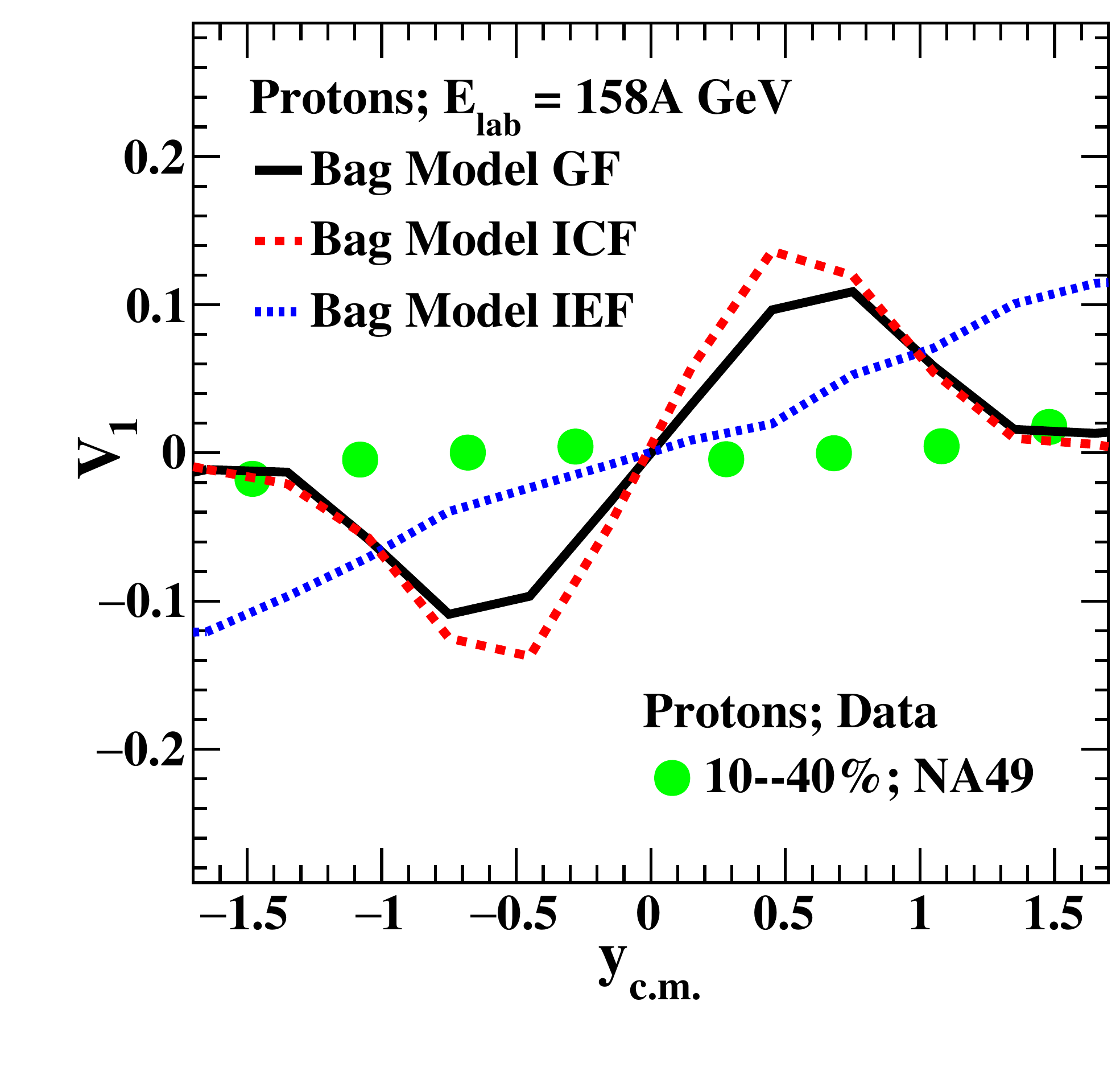}\\
\includegraphics[scale=0.295]{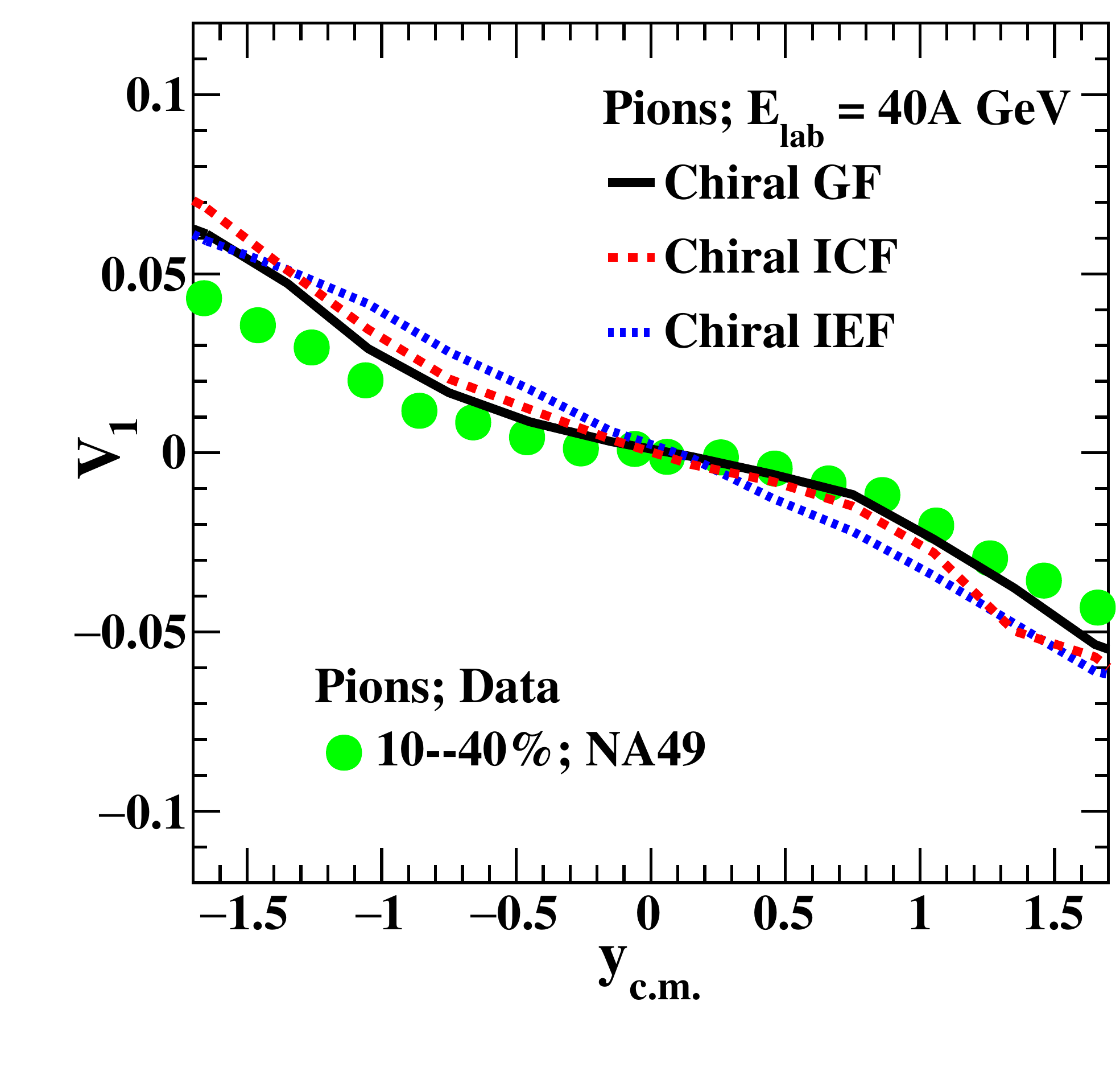}
\includegraphics[scale=0.295]{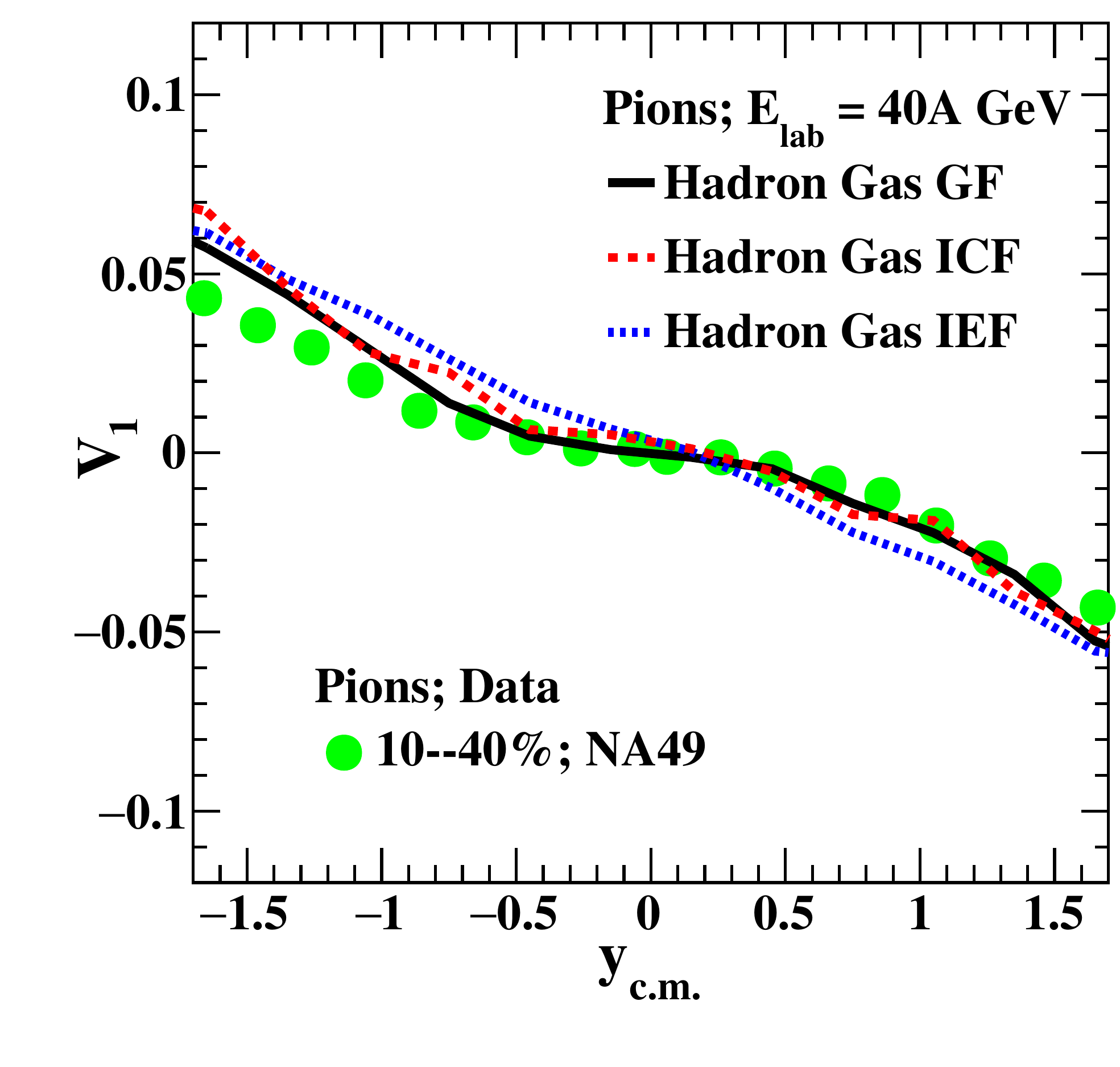}
\includegraphics[scale=0.295]{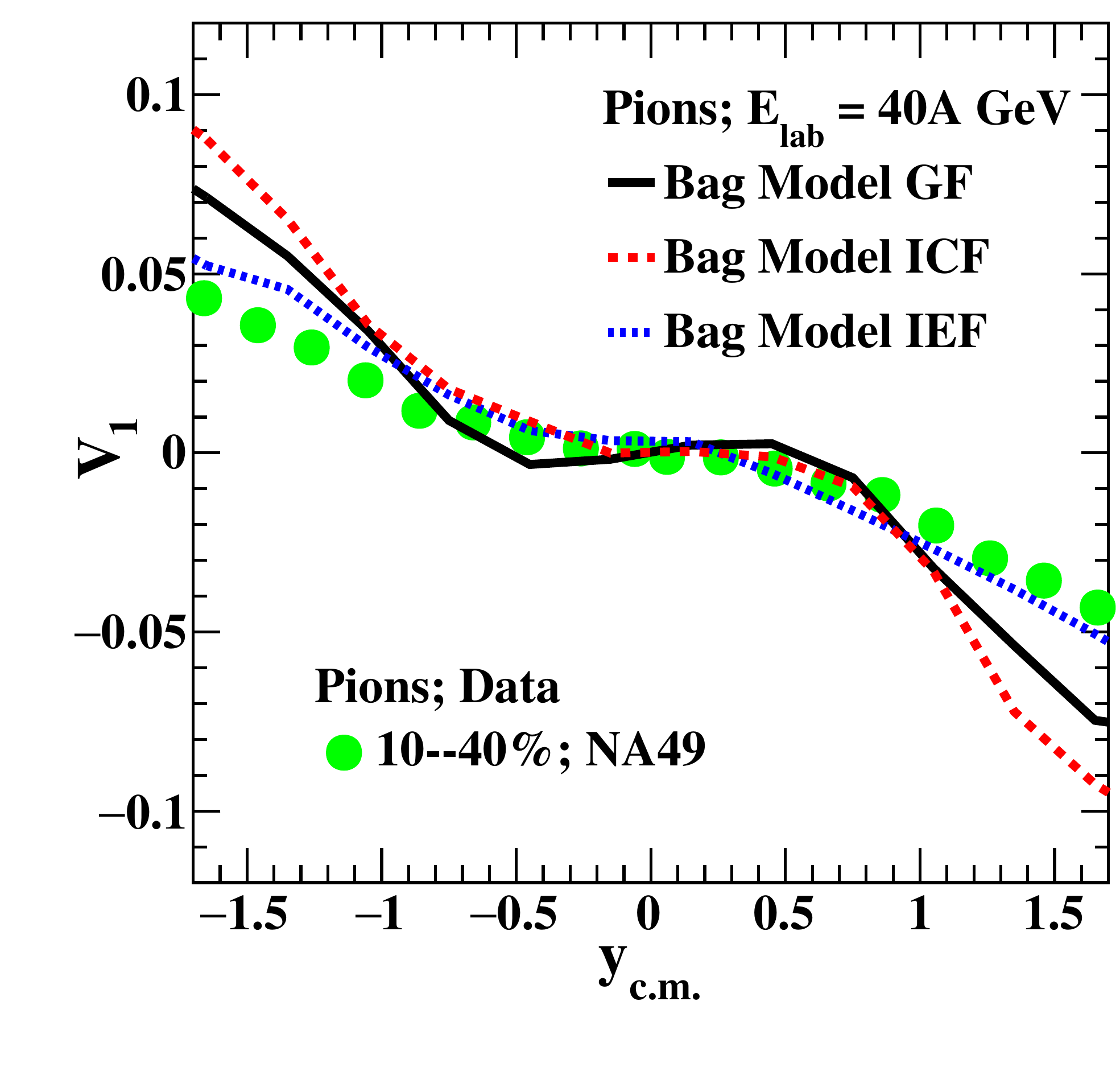}\\
\includegraphics[scale=0.295]{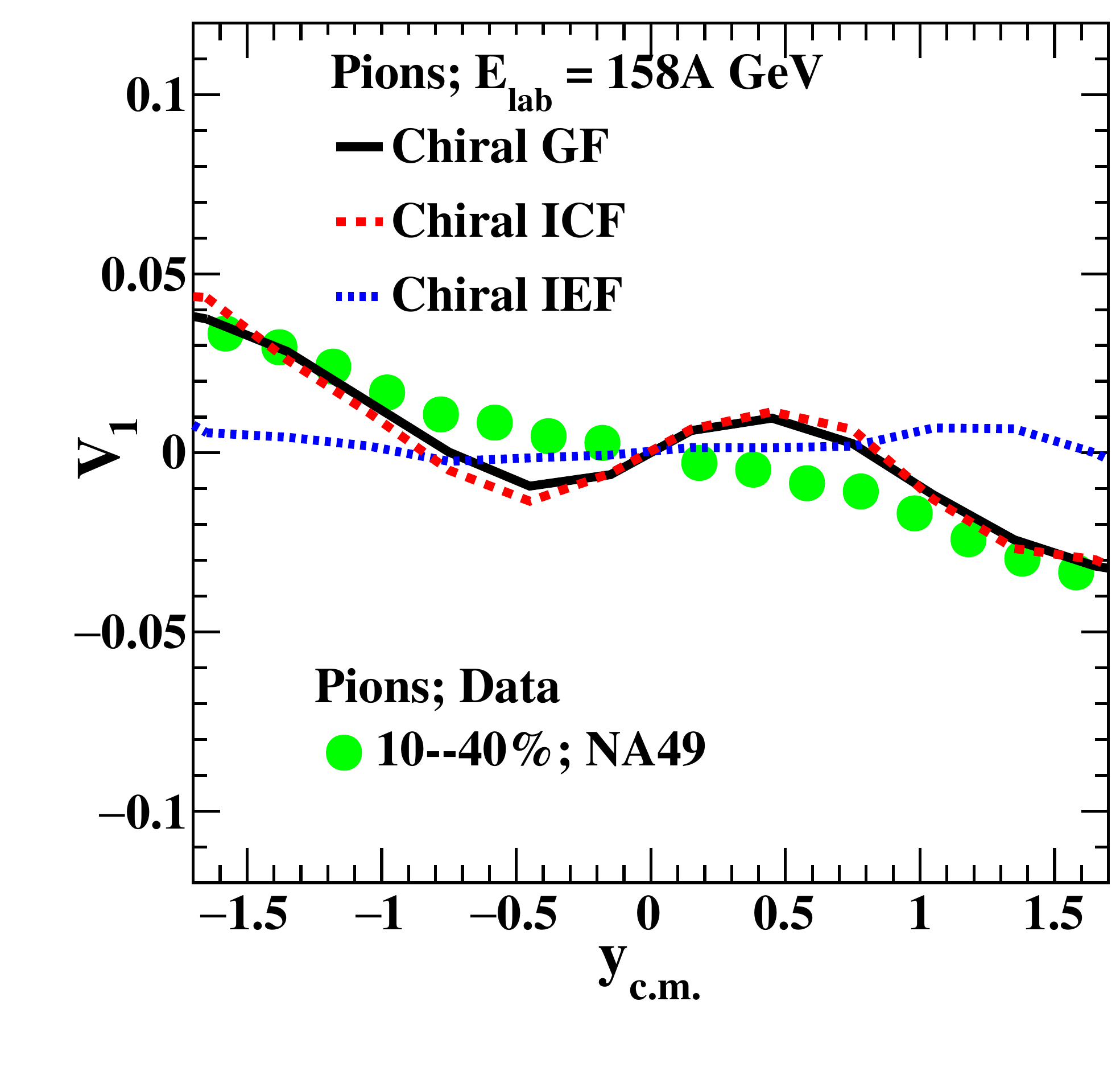}
\includegraphics[scale=0.295]{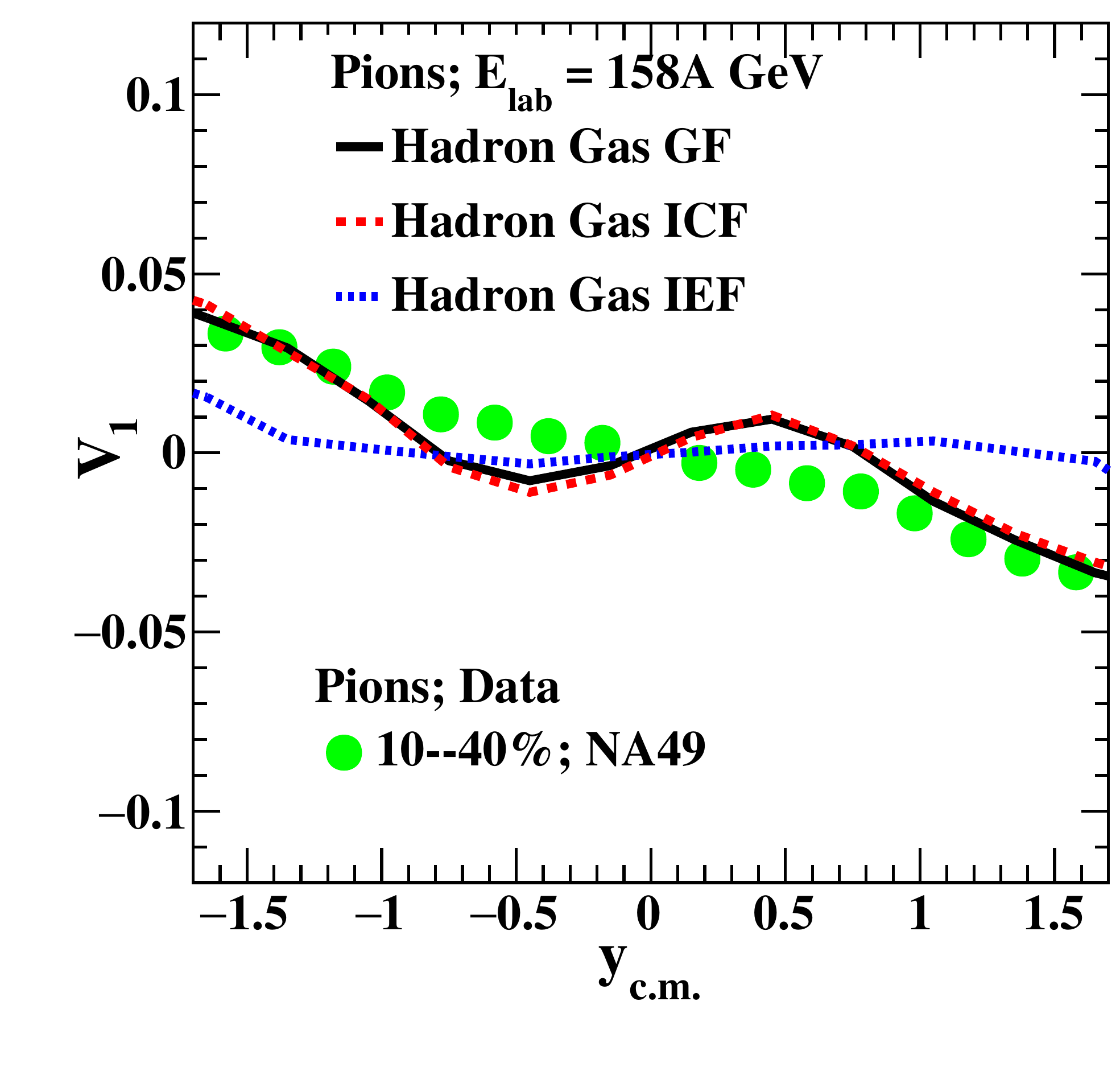}
\includegraphics[scale=0.295]{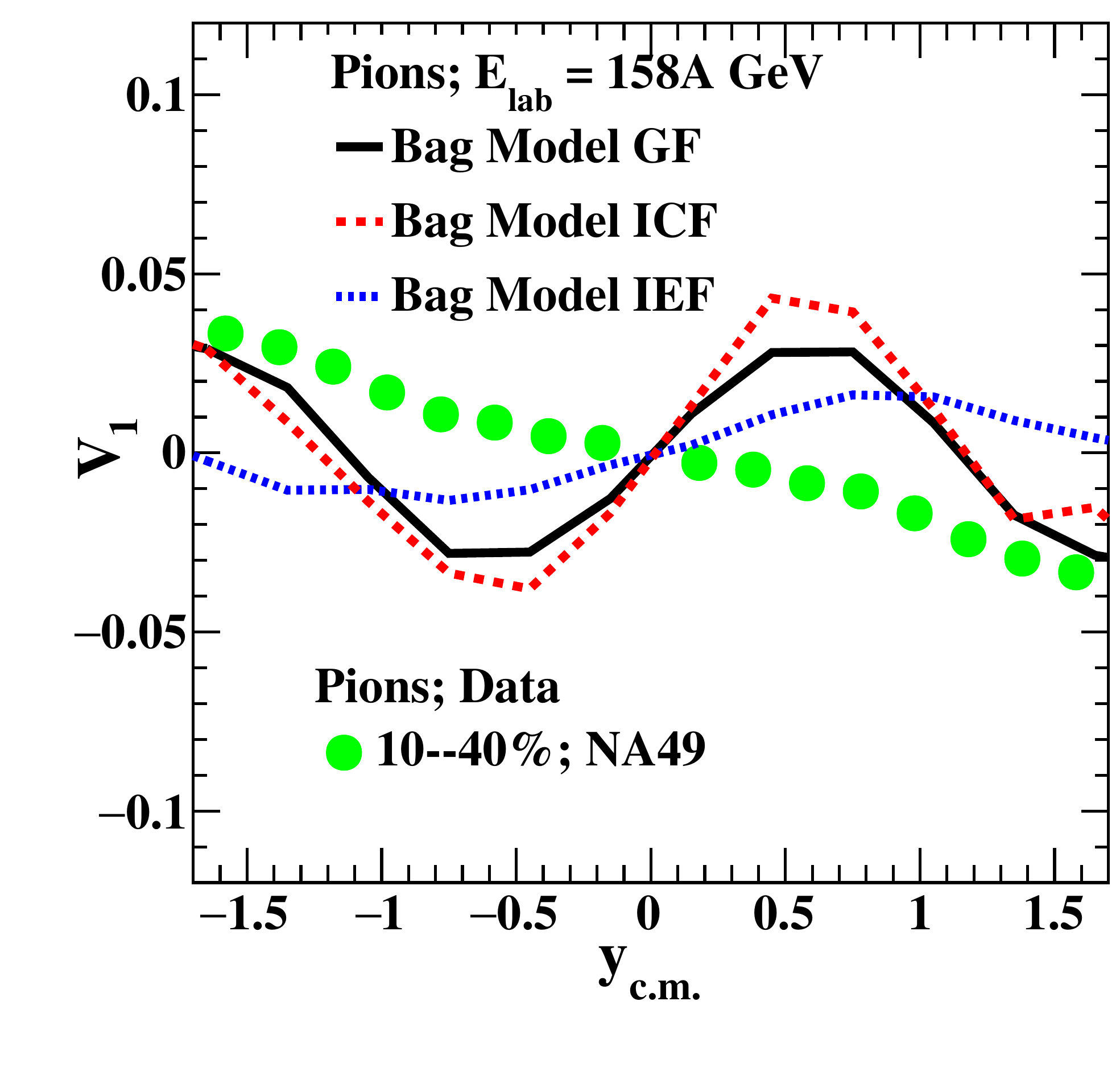}\\

\caption{(Color online) Comparison of directed flow of protons (upper two panels) and pions (bottom two panels) as a function of rapidity with experimental measurements at 40A and 158A GeV~\cite{Alt:2003ab} at SPS for different EoS and particlization modes of UrQMD with measured directed flow for non-central (b = 5-9 fm) Au-Au collisions.}
\label{figprotons}
\end{figure*}

\section*{II. Model description}
The working principle of the ultrarelativistic quantum molecular dynamics (UrQMD) model can be found in Refs~\cite{Bass:1998ca,Bleicher:1999xi,Petersen:2008dd}. Simulation of nucleus-nucleus collisions is the objective of the UrQMD model. Woods-Saxon profile and Fermi gas model are employed to initialize the target and projectile nuclei in coordinate and momentum space. With the experimental inputs, namely, cross-sections and decay widths, the models incorporate the collisions as interactions among hadrons and resonances at low energies and string excitation and fragmentation at higher energies.

In the hybrid version of the UrQMD model, pure transport calculations are coupled with hydrodynamics calculations using the SHASTA algorithm~\cite{Rischke:1995ir,Rischke:1995mt} to model the intermediate hot and dense stage of the heavy-ion collision. 
As soon as the two Lorentz-contracted nuclei cross each other, which corresponds to the time that ensures all baryon scatterings and energy deposition have taken place, the hydrodynamics is switched on. This time act as a lower bound for thermalization time. Right after this, the spectators are sent into the cascade, and participant particles, which are "point-like" in nature, are mapped to a hydrodynamical grid. 

Now, the time is for hydrodynamical evolution, where the equation-of-state (EoS) is an essential input. There are three EoS available in the public version of the UrQMD hybrid model. The first one is the hadron gas (HG) EoS~\cite{Zschiesche:2002zr} which is non-interacting gas of hadrons expressed by grand canonical ensemble and does not have any phase transition. As the underlying degrees of freedom are the same, the hydrodynamical and pure transport approaches can be compared on equal footing. The next EoS is chiral + deconfinement EoS~\cite{Steinheimer:2011ea} which incorporates both chiral and deconfinement phase transitions. The latter is of cross-over type in nature for all finite net baryon densities. In addition to this, this EoS has partonic degrees of freedom. Moreover, deconfinement transition occurs through quarks and Polyakov potential, whereas hadronic interaction administers chiral phase transition. Note that the partonic degrees of freedom only show up when the temperature is high enough. It has been seen that this EoS agrees well with the lattice QCD simulations at vanishing baryon chemical potential. The final available EoS is the bag model (BG)~\cite{Rischke:1995mt} which is a combination of the standard MIT bag model and an improved version of the $\sigma-\omega$ model. The former is utilized for the QGP phase, whereas the latter is employed in case of the hadronic phase. This EoS is available with an inbuilt first-order phase transition, during which Gibbs' conditions are used to match both hadronic and partonic phases for equilibrium.

As time goes on, the system starts to become diluted, and fluid dynamical evolution would no longer be applicable. Therefore, the fluid description can be changed to particle description using Cooper-Frye formalism, known as "Particlization". One of the crucial steps here is the determination of transition hypersurface. In the hybrid UrQMD model, depending on the type of hypersurface and switching criteria, three different particlization scenarios are available for fluid to particle transition. The default scenario is known as the gradual particlization scenario (GF), in which the slices of 0.2 fm thick undergo particlization when the energy density in all cells of that slice falls below five times the ground state energy density $\epsilon_0$. So, the particlization is performed slice by slice. The following scenario is isochronous particlization (ICF), where the particlization takes place at the same time as soon as the energy density in all cells falls below the critical value (5$\epsilon_{0}$). In these two cases, the hypersurface is isochronous in nature. In the latter case, it is possible that the parts of the system may have become dilute as the transition can only be done once all cells have energy density below a specific critical value. Therefore, the application of fluid dynamics would be questionable in this case. 

The last one is the iso-energy density particlization scenario (IEF), where the iso-energy density hypersurface is constructed numerically with the help of Cornelius routine~\cite{Huovinen:2012is}. After that, the hydrodynamical fields are mapped to particles on this hypersurface using Cooper-Frye formalism once the energy density in all cells reaches below the critical value. As discussed in Ref.~\cite{Huovinen:2012is}, this scenario is suitable for dealing with event-by-event heavy-ion collisions analysis. The sampling algorithm described there is very flexible in event-by-event calculations where the initial states of hydrodynamics change widely.

\section*{III. Results and Discussion}
In this section, we present the outcomes of our investigation of the dependence of various experimental observables on the available particlization models in the UrQMD hybrid model. The study is performed using noncentral Au + Au collisions for various beam energies in the range 1A--158A GeV, and the results are compared with experimental measurements in similar kinematic regions wherever available. To begin, we investigate various flow coefficients for a range of beam energies using different particlization modes. Then we see the effect of particlization prescriptions on particle production by looking at strange to non-strange ratio, baryon to mason ratio, and more. Then we look at the net proton rapidity spectra for different particlization modes in combination with different EoS.

\subsection*{A. Anisotropic flow coefficients}

In regards to the paper, the study of anisotropic coefficients for particlization models is essential. The reason being the choice of hypersurface and switching criteria allows the fluid dynamic evolution to conclude at different times. Particlization may affect the flow of the species as it will spend more or less time in evolution. Directed flow is sensitive to longitudinal dynamics among various anisotropic flow coefficients. So we estimate the directed flow of pions and protons as a function of rapidity for various particlization scenarios and equations of state as shown in Fig.~\ref{figprotons}. The results are compared with the experimental data from NA49 experiment~\cite{Alt:2003ab} at SPS for similar centrality regions (10--40\%) at 40 and 158A GeV. In the case of pions for 40A GeV, all particlization scenarios favor the experimental result and do not show much difference. If we look at pions and protons for higher beam energy at 158A GeV, the iso-energy particlization shows more favorable results than the other two particlization modes. One interesting point to note here is that the so-called "wiggle" structure is less pronounced in the cases where IEF is employed compared to the other two scenarios.

\begin{figure*}
\begin{center}
\includegraphics[scale=0.35]{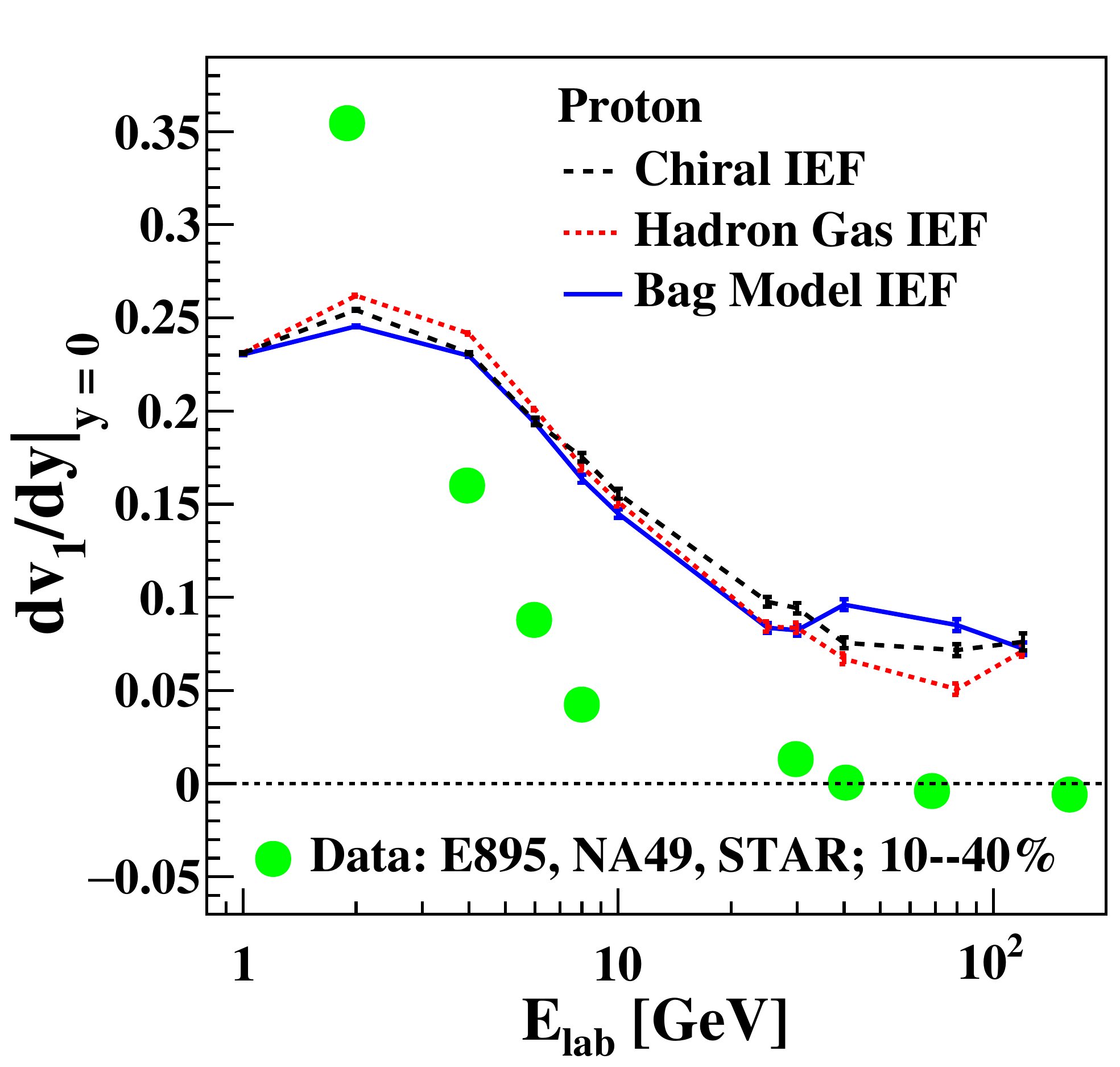}
\includegraphics[scale=0.35]{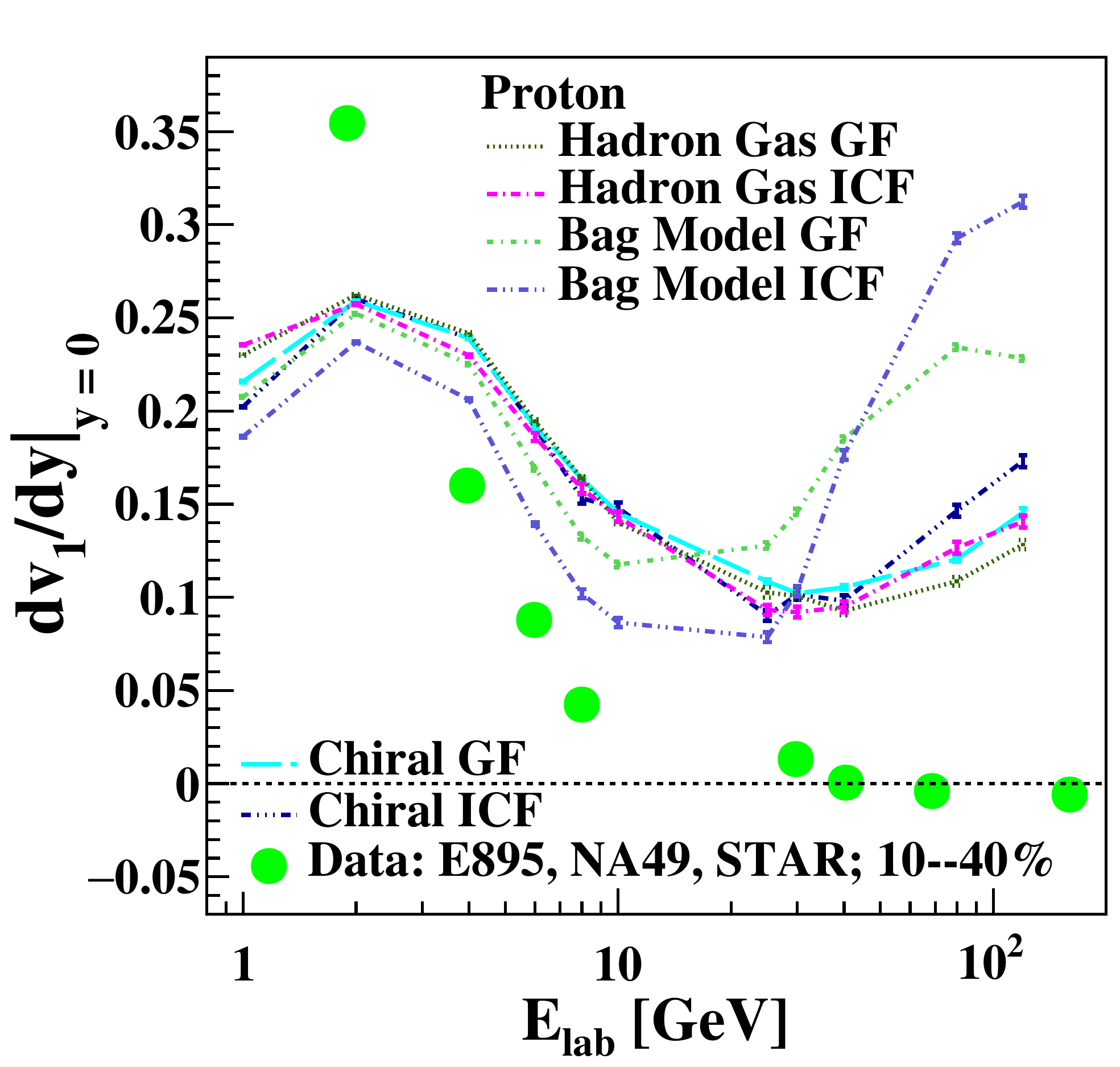}
\end{center}
\caption{(Color online) Comparison of slope of the directed flow of protons as a function of beam energy at midrapidity for different EoS and particlization modes of UrQMD for non-central (b = 5-9 fm corresponds to approximately 10-40\% central) Au-Au collisions with E895~\cite{Liu:2000am} and STAR~\cite{Adamczyk:2014ipa} experimental measurements in Au-Au collisions and with NA49~\cite{Alt:2003ab} experimental measurements in Pb-Pb collisions.}
\label{fig2}
\end{figure*}

We choose the slope of the directed flow of protons as our next observable for investigation, which is obtained by first-order polynomial fitting of differential direct flow $[v_{1}(y)]$ at the mid-rapidity. The slope of directed flow at mid-rapidity is an interesting observable and contains insights into the medium properties. In Fig.~\ref{fig2}, the results for proton with $p_{\rm T} < 2$ GeV/c are compared with the measurements of E895~\cite{Liu:2000am}, NA49~\cite{Alt:2003ab}, and STAR~\cite{Adamczyk:2014ipa} experiments.

\begin{figure*}
\includegraphics[scale=0.295]{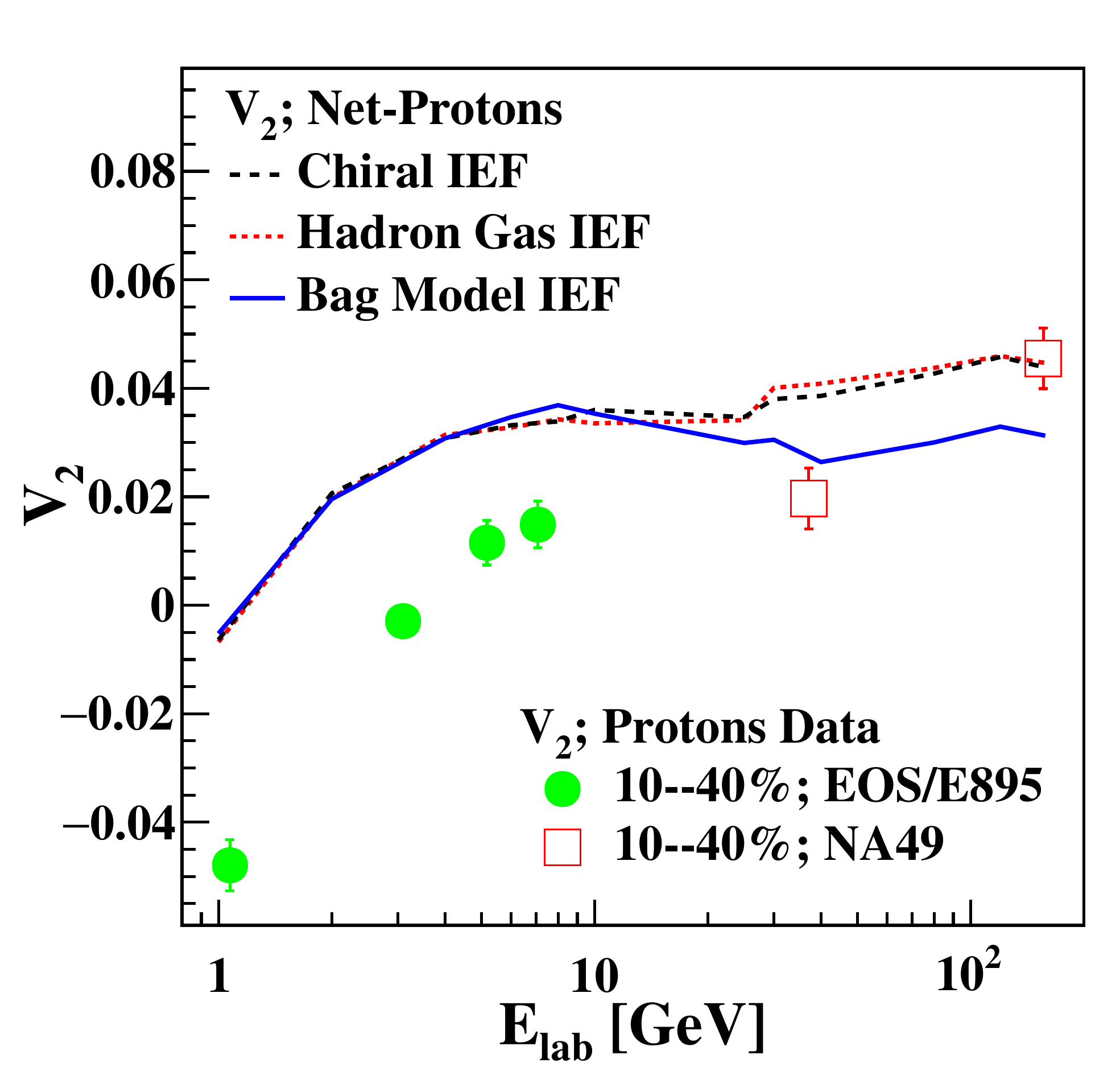}
\includegraphics[scale=0.295]{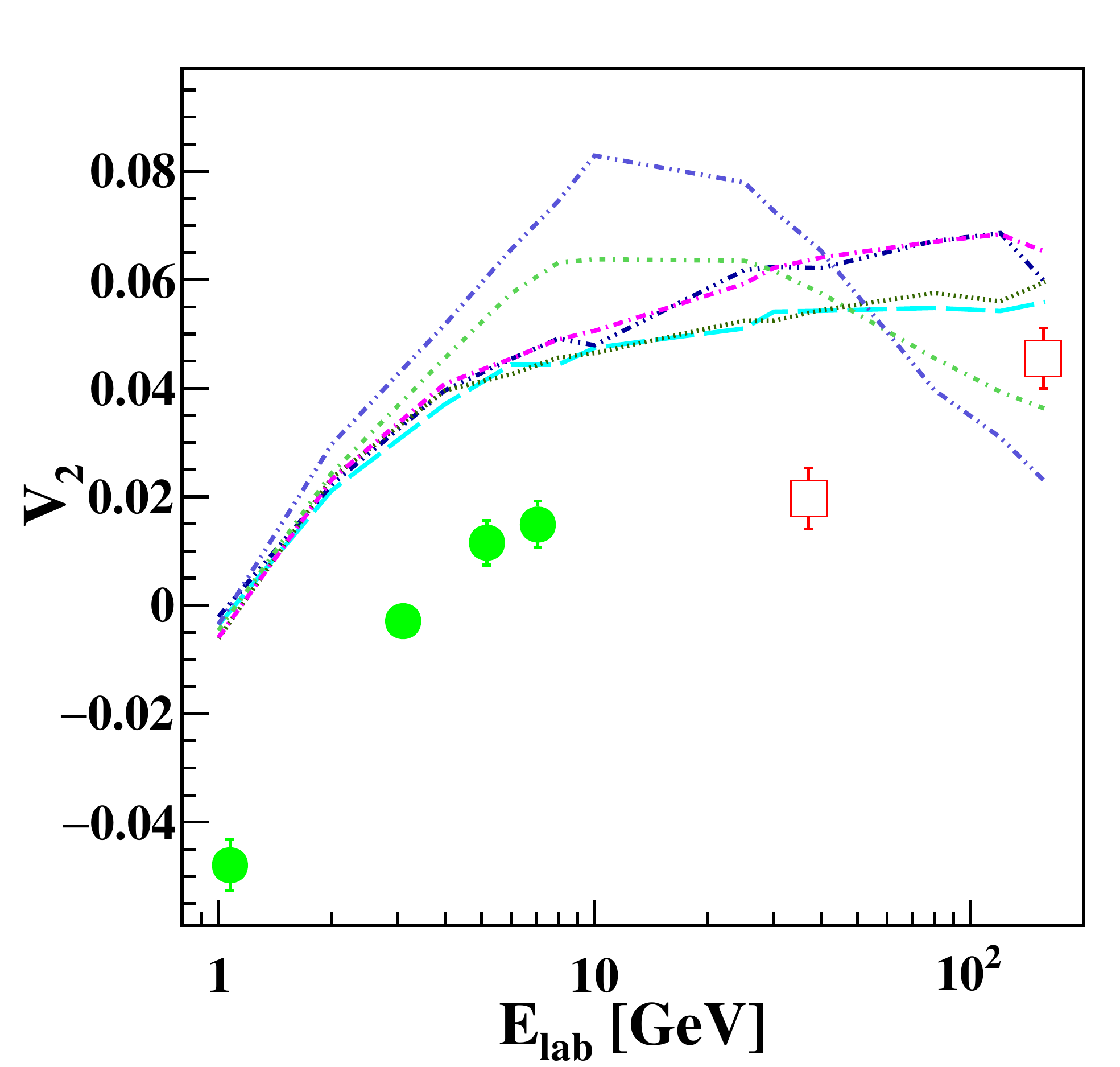}
\includegraphics[scale=0.295]{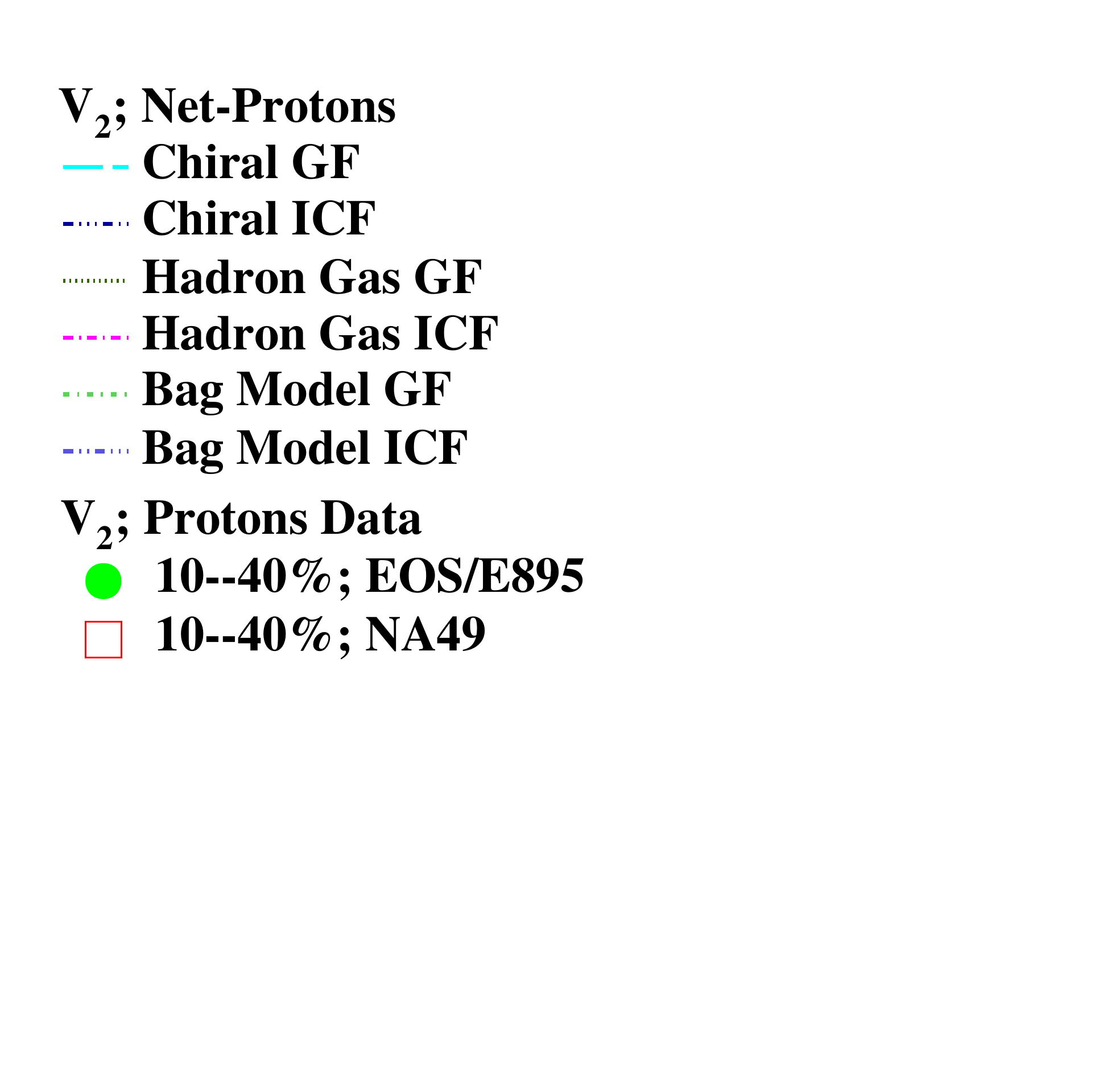}\\
\includegraphics[scale=0.295]{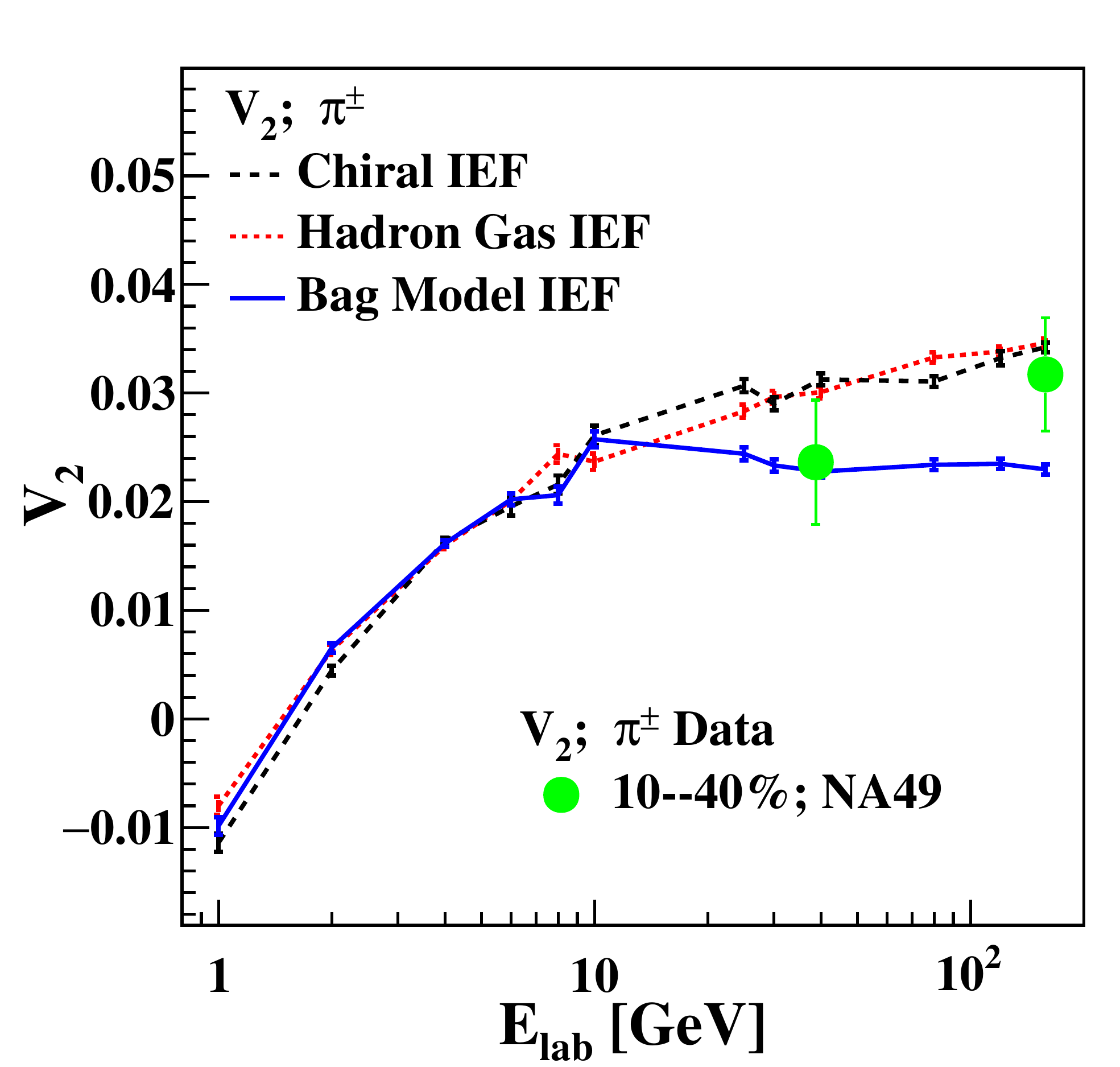}
\includegraphics[scale=0.295]{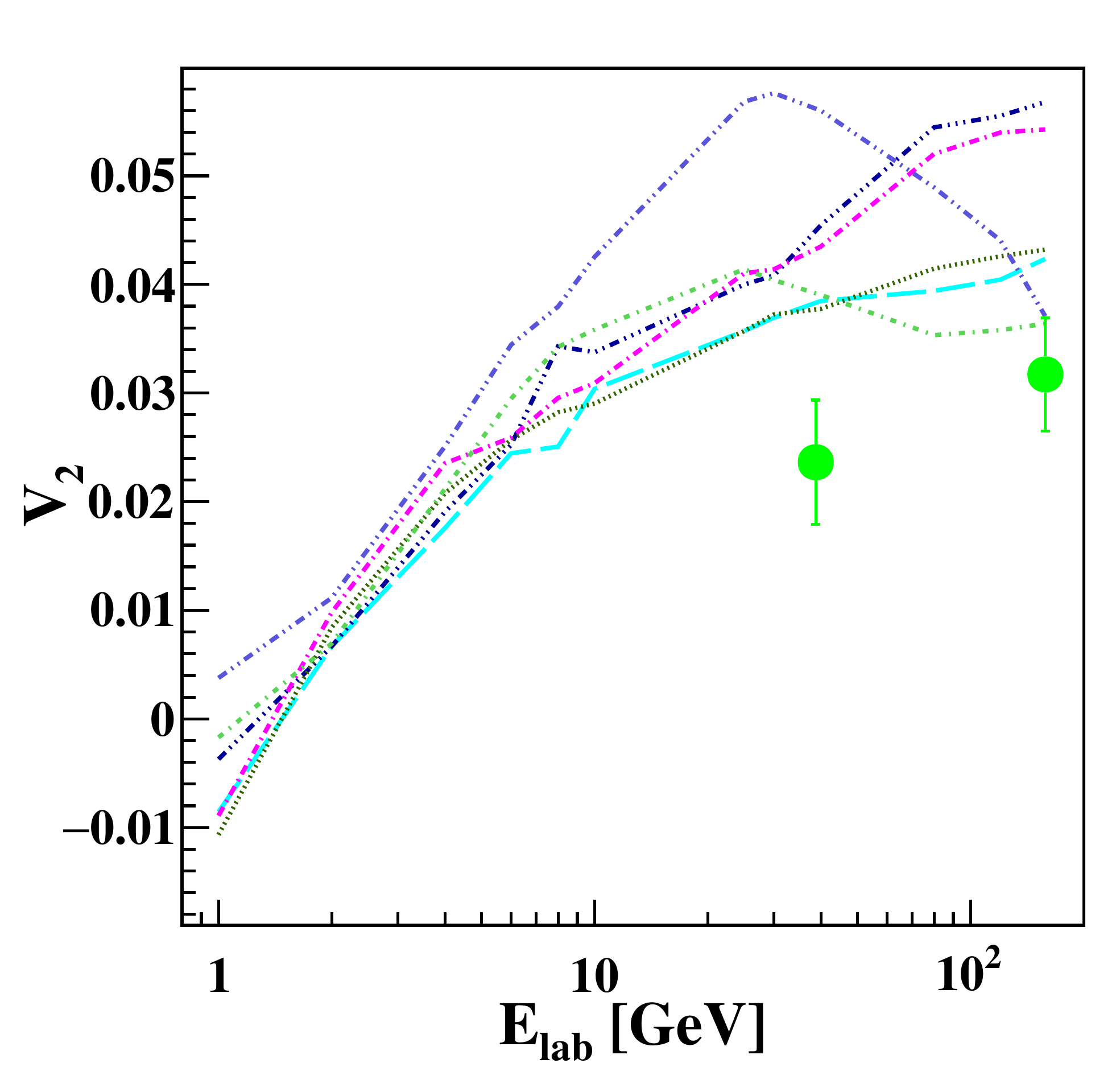}
\includegraphics[scale=0.295]{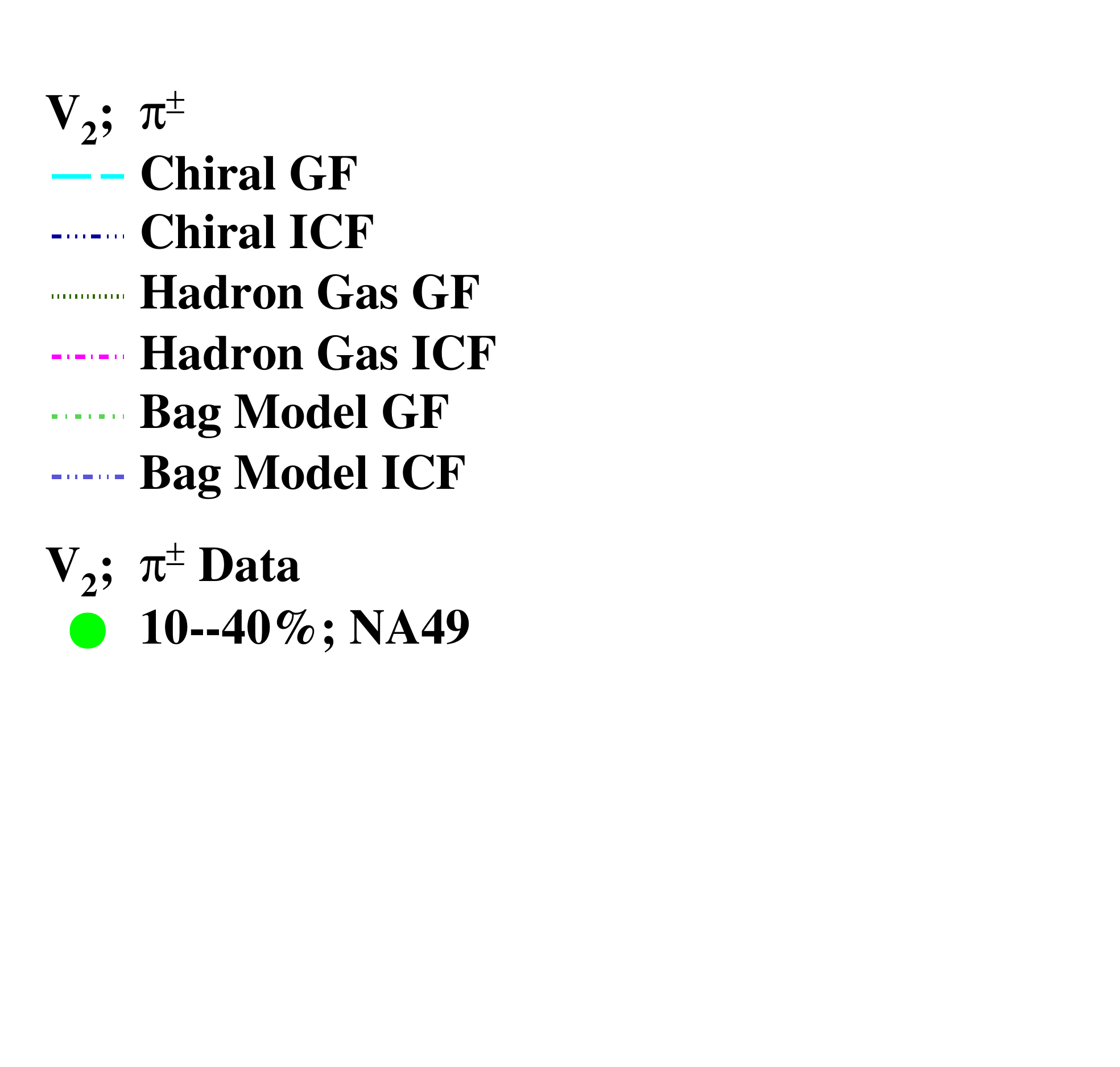}\\
\caption{(Color online) $p_{\rm T}$ integrated elliptic flow $v_{2}$ of net protons, and pions as a function of beam energy at midrapidity (-0.5 $<$ $y_{c.m.}$ $<$ 0.5) for different EoS and particlization modes of UrQMD for non-central (b = 5-9 fm corresponds to approximately 10-40\% central) Au-Au collisions. $v_{2}$ of net protons and pions for $p_{\rm T}$ $<$ 2 GeV/c are compared with available E895 and NA49 experimental measurements~\cite{Pinkenburg:1999ya,Alt:2003ab} respectively in the investigated beam energy range in Au-Au and Pb-Pb collisions.}
\label{fig3}
\end{figure*}

\begin{figure*}
\includegraphics[scale=0.295]{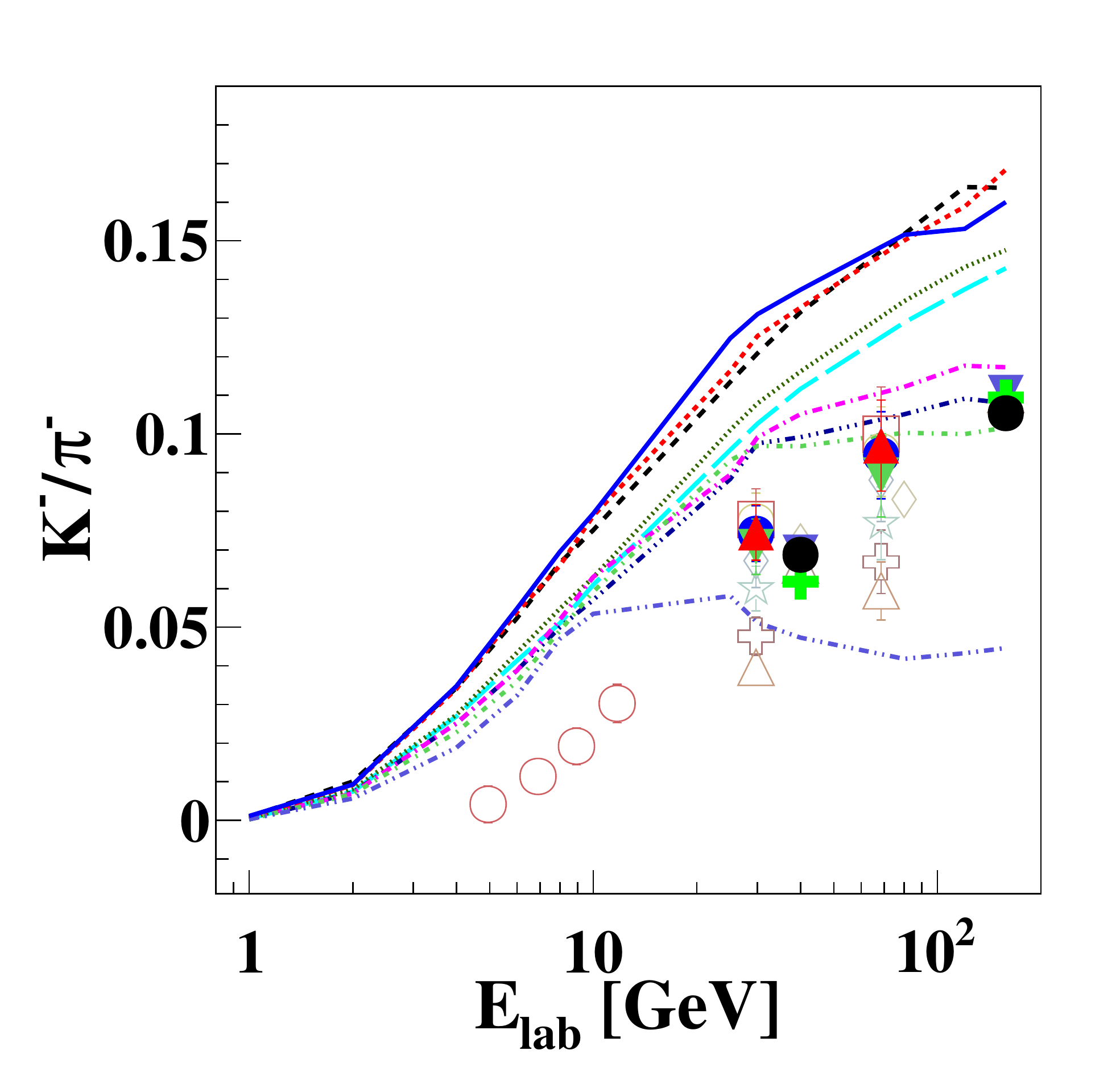}
\includegraphics[scale=0.295]{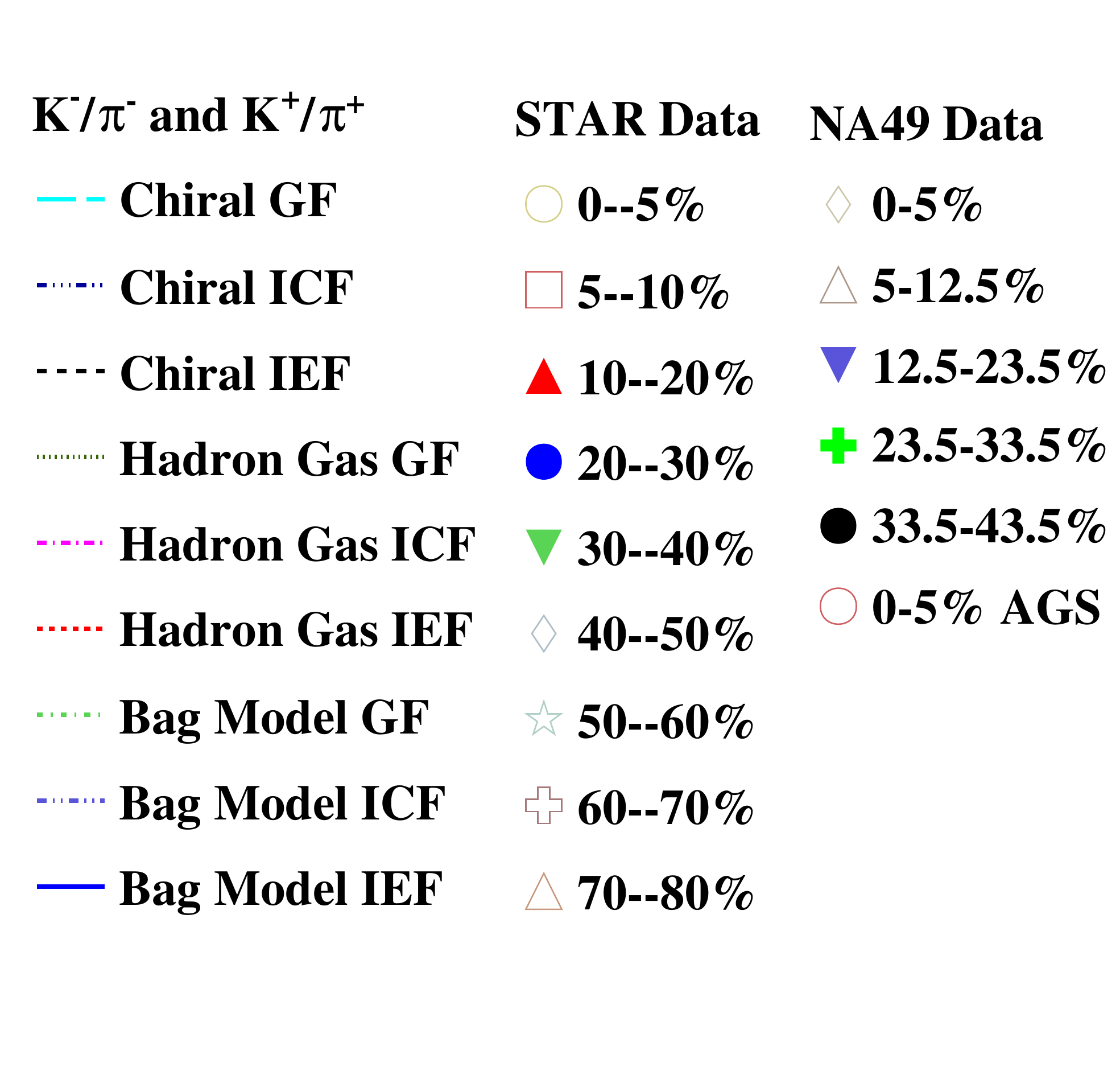}
\includegraphics[scale=0.295]{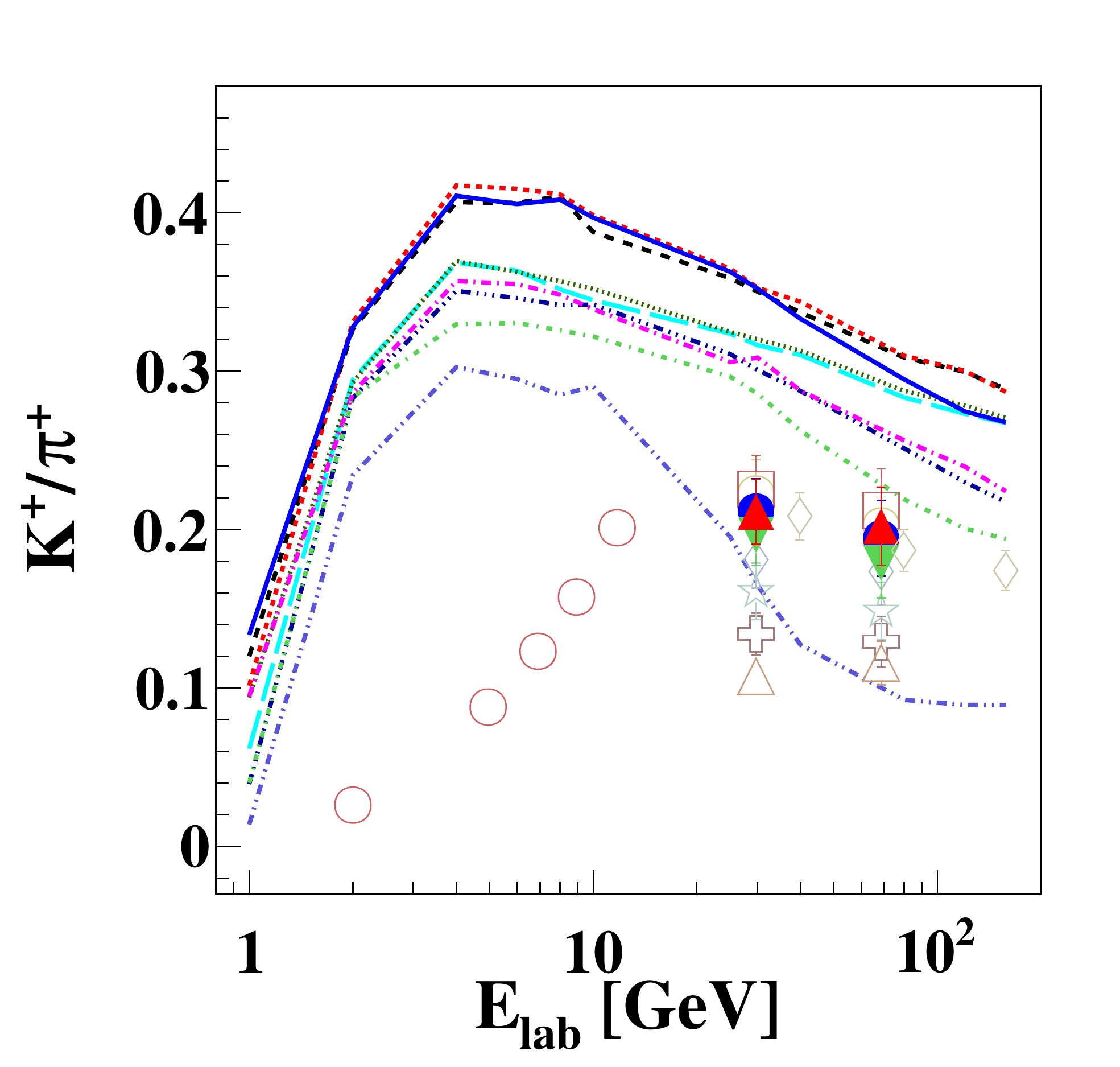}\\
\caption{(Color online) $\rm K^{-}$ to $\pi^{-}$, and $\rm K^{+}$ to $\pi^{+}$ ratio as a function of beam energy for different EoS and particlization modes of UrQMD for non-central (b = 5-9 fm corresponds to approximately 10-40\% central) Au-Au collisions and their comparison with AGS~\cite{E866:1999ktz}, NA49~\cite{NA49:2002pzu,Alt:2006dk}, and STAR experimental measurements~\cite{Adamczyk:2017iwn} in Au-Au, Pb-Pb, and Au-Au collisions respectively for all available centralities. Vertical bars on the data denote statistical uncertainties.}
\label{fig4}
\end{figure*}

\begin{figure*}
\includegraphics[scale=0.295]{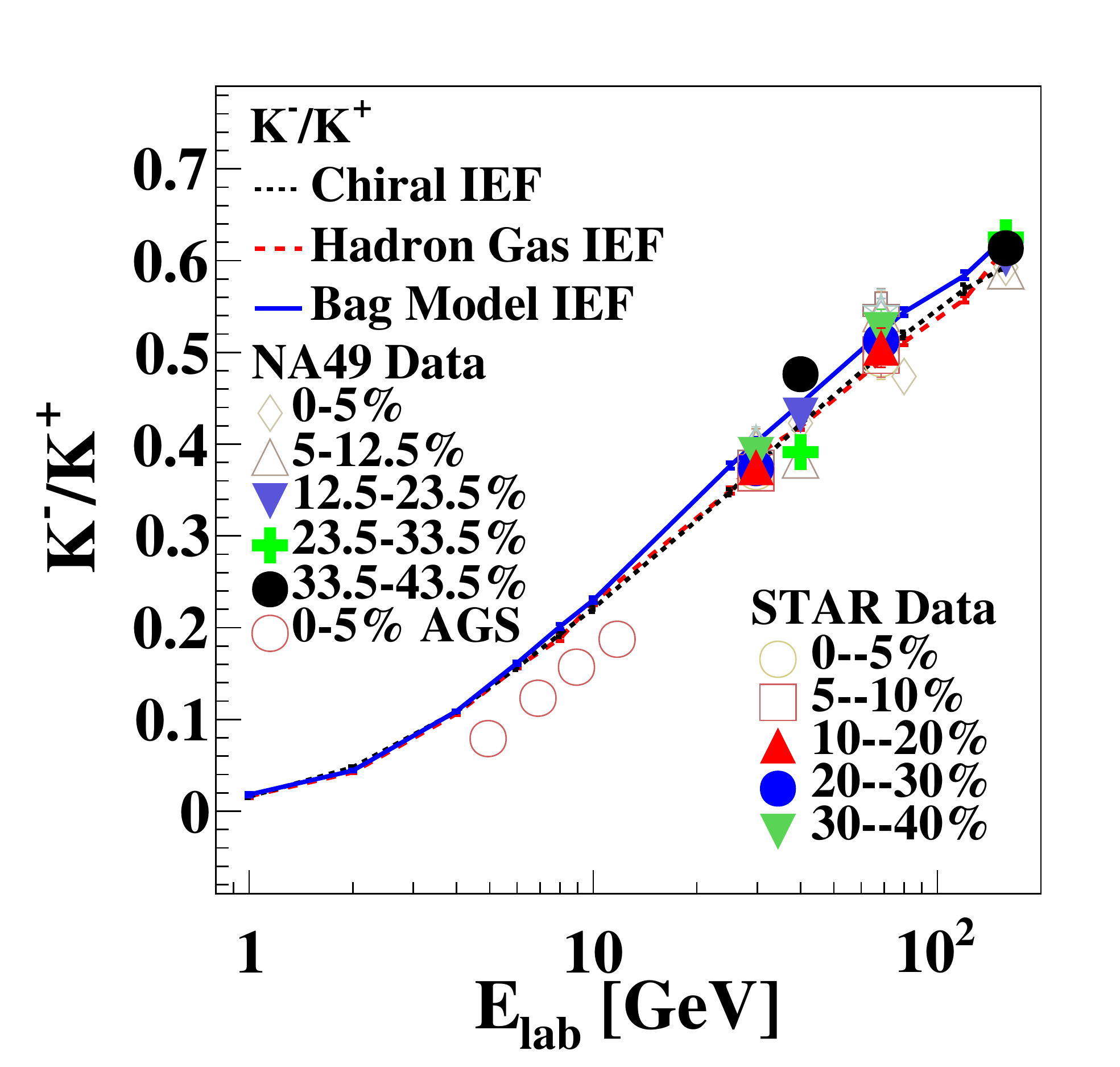}
\includegraphics[scale=0.295]{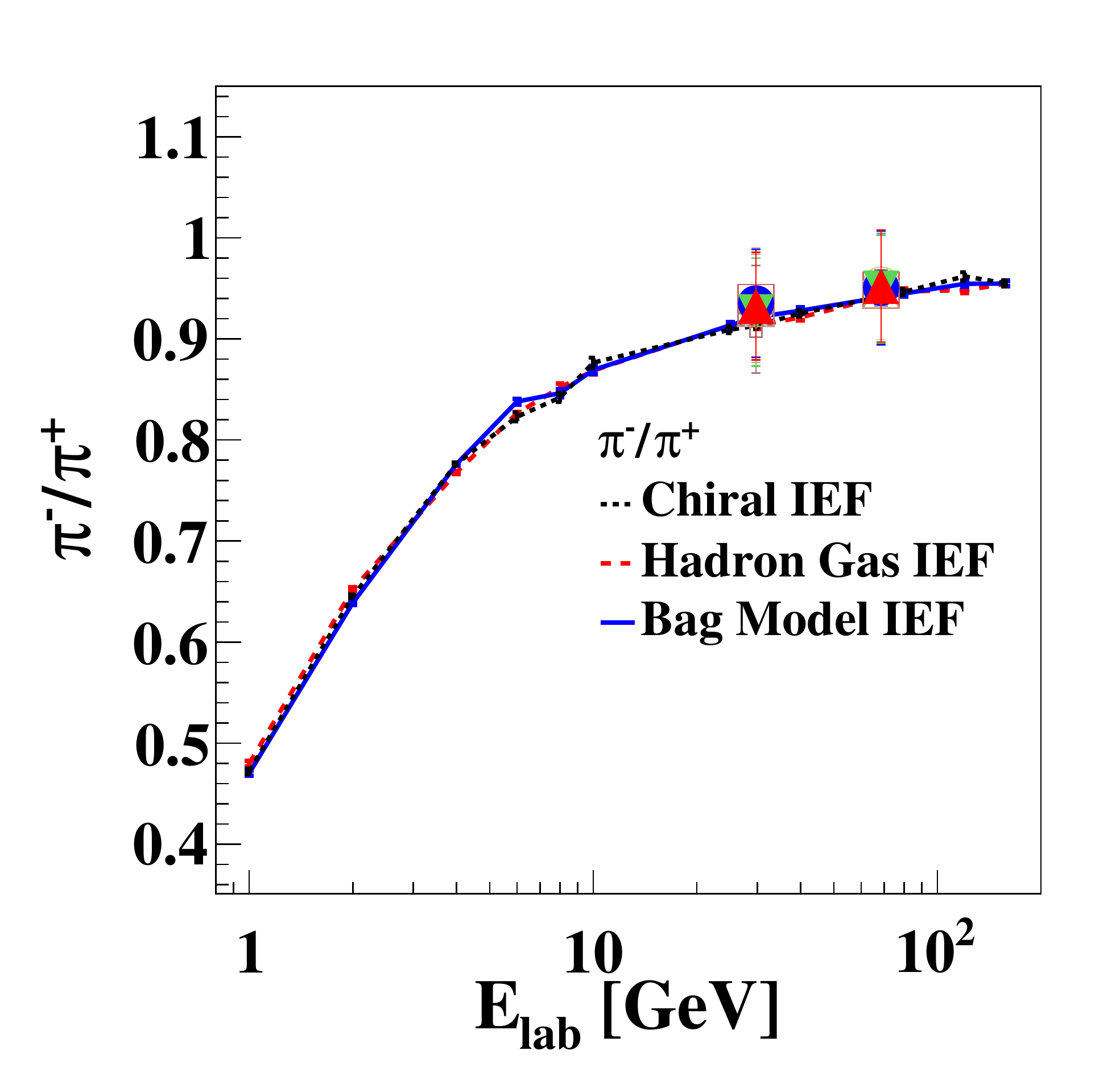}
\includegraphics[scale=0.295]{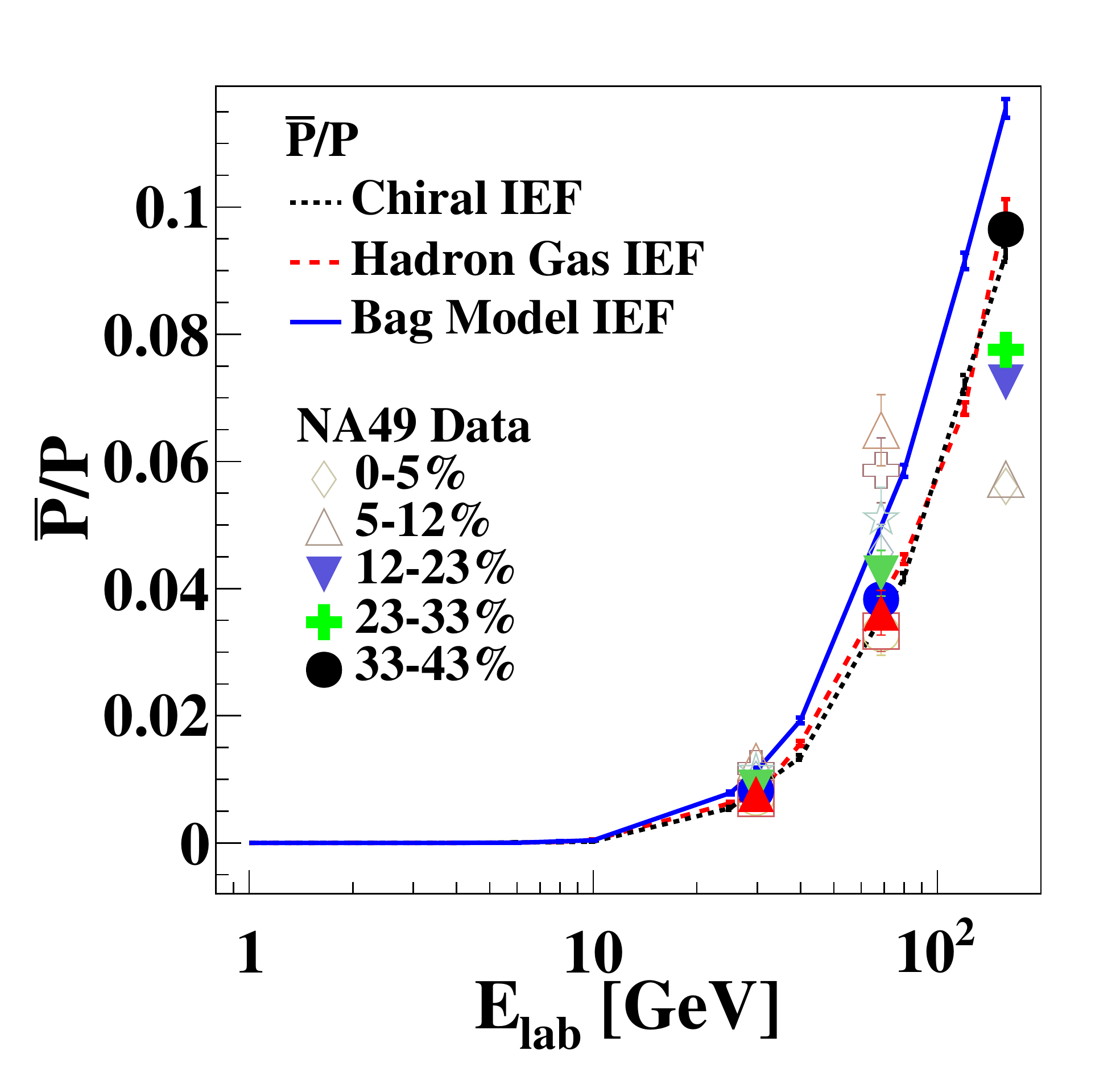}\\
\includegraphics[scale=0.295]{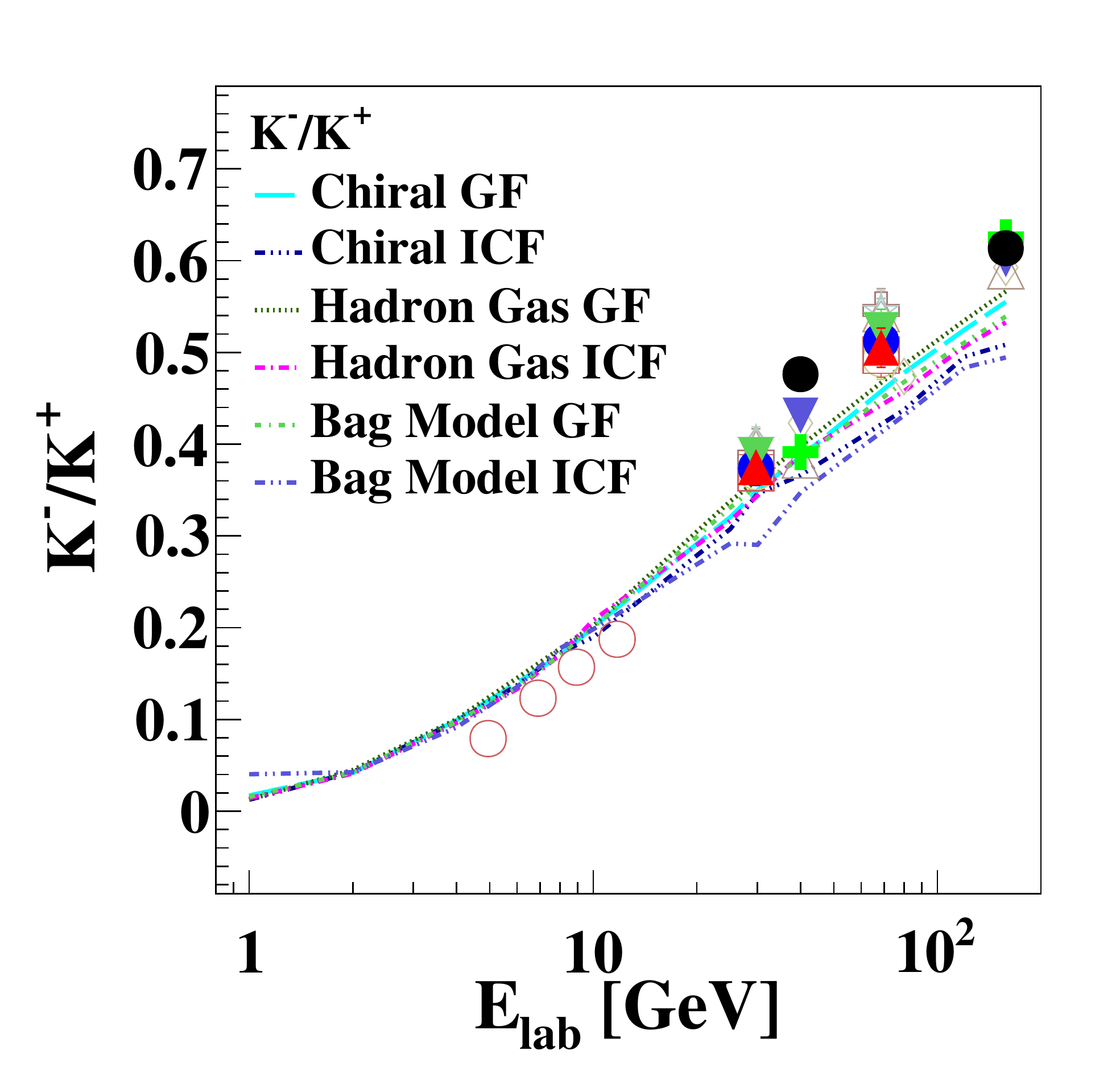}
\includegraphics[scale=0.295]{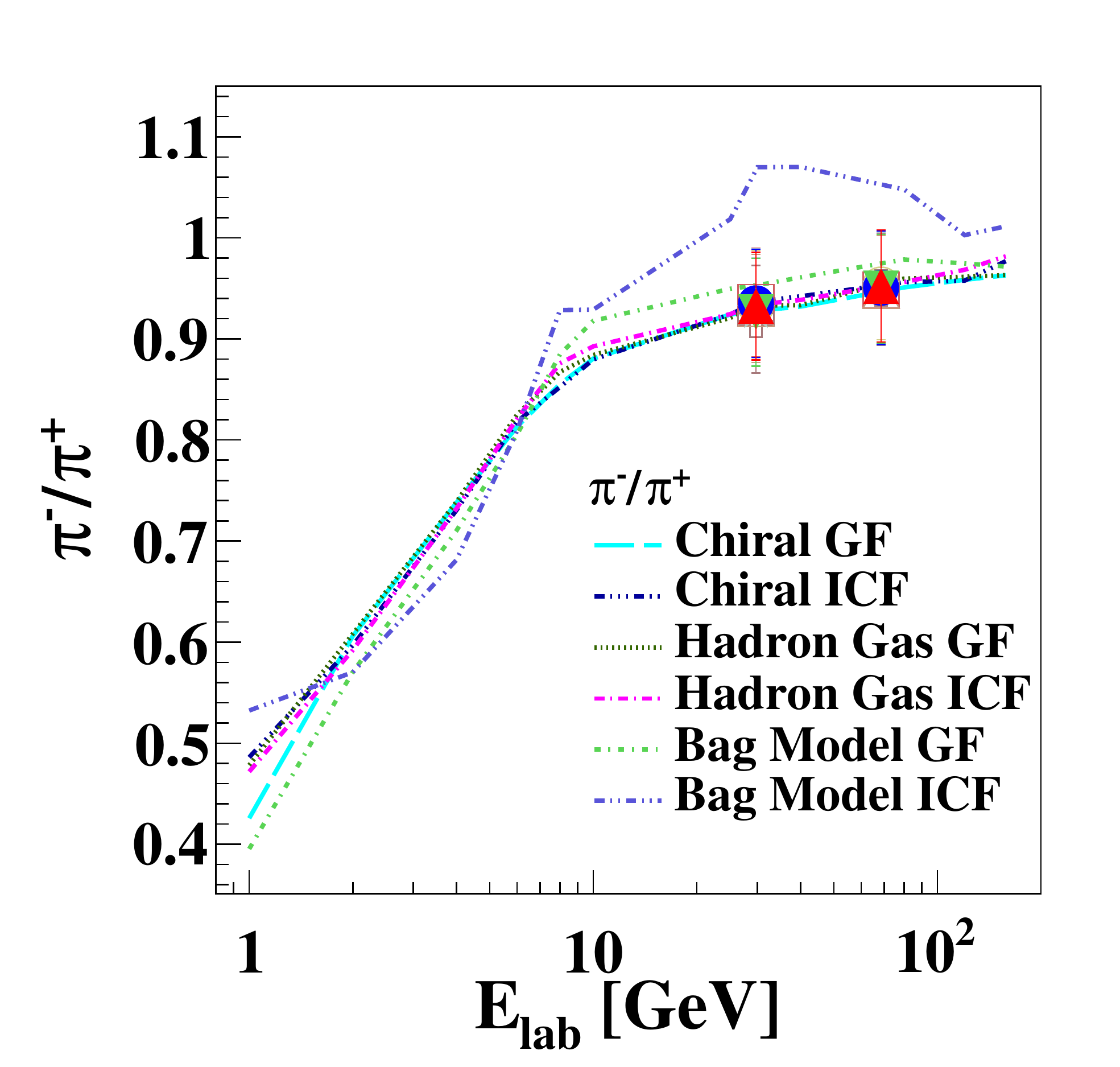}
\includegraphics[scale=0.295]{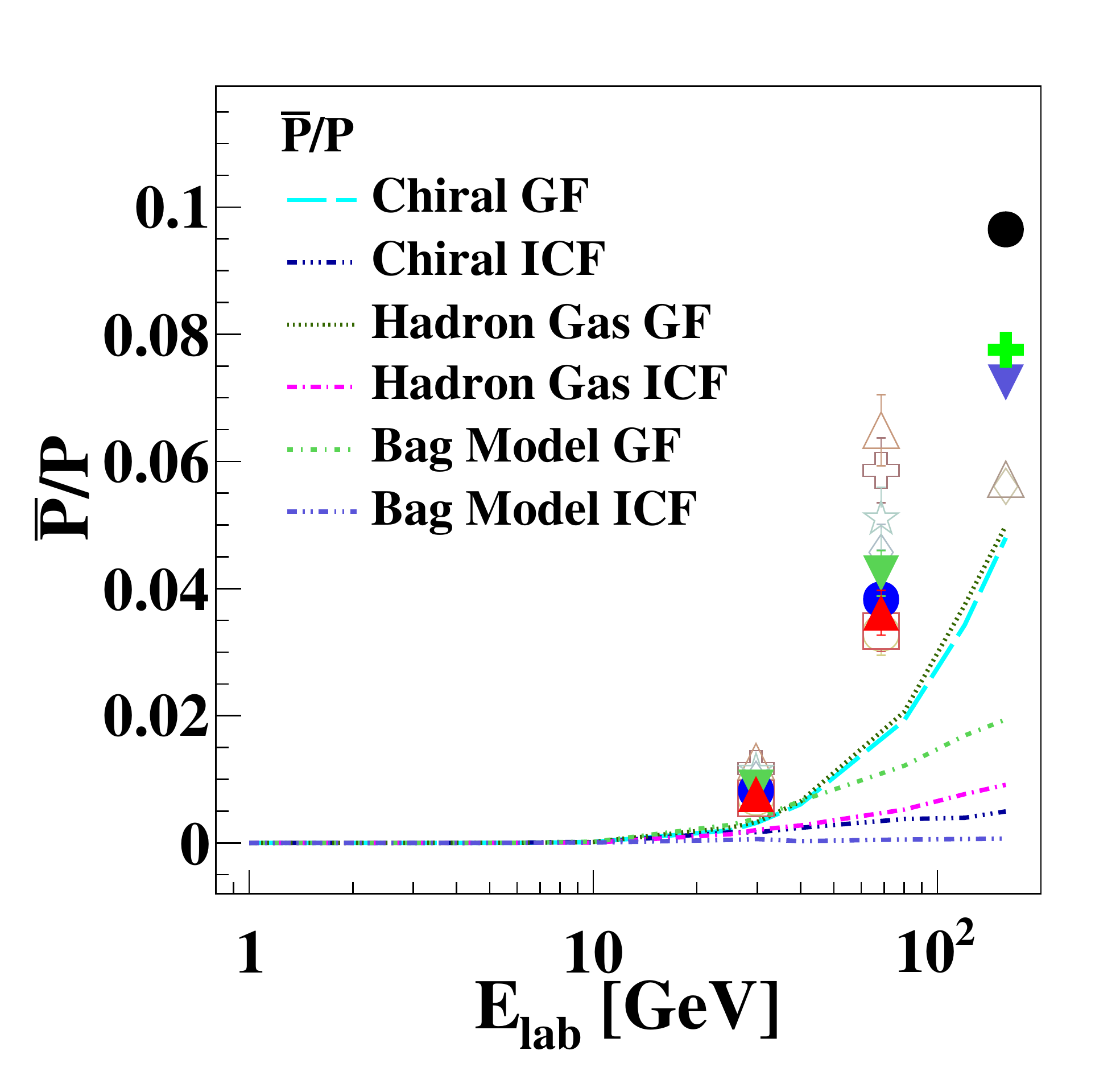}\\
\caption{(Color online) $\rm K^{-}$ to $\rm K^{+}$, $\pi^{-}$ to $\pi^{+}$ and anti-proton to proton ratio as a function of beam energy for different EoS and particlization modes of UrQMD for non-central (b = 5-9 fm corresponds to approximately 10-40\% central) Au-Au collisions and their comparison with AGS~\cite{E866:2000dog}, NA49~\cite{Alt:2006dk} and STAR experimental measurements~\cite{Adamczyk:2017iwn} in Au-Au, Pb-Pb and Au-Au collisions respectively for all available centralities. Vertical bars on the data denote statistical uncertainties.}
\label{fig5}
\end{figure*}

It can be seen that the slope of the directed flow of protons does not show any sensitivity to the underlying degrees of freedom below 20A-30A GeV for all particlization scenarios. This output agrees with the claims made in one of our works~\cite{Kundu:2021afz} which is due to short lifetime of hydrodynamical phase. In our previous study, we observed a splitting among the comparison of various EoS~\cite{Kundu:2021afz}; here, similar splitting is exhibited around 20--30 AGeV. This suggests that various particlization models and EoS behave differently above this energy range, perhaps indicating the threshold for the possible onset of deconfinement. However, a detailed study in this direction is needed to make any firm statement on this.

Moving on to elliptic flow ($v_{2}$), which was estimated for protons and pions and studied as a function of beam energies for different EoS and particlization models as shown in Fig.~\ref{fig3}. The observable is compared with measurements from EOS/E895~\cite{Pinkenburg:1999ya} and NA49~\cite{Alt:2003ab} experiments. Similar to the previous two observables, $v_{2}$ in the case of IEF scenario shows better agreement with experimental measurements at high beam energies, whereas $v_{2}$ for ICF case diverges from the data even more than default GF scenario. Neither particlization model agrees with measured $v_{2}$ at low beam energies. One of the observations is that $v_{2}$ using the bag model shows non-monotonous behavior irrespective of the particlization modes for both species. It seems that because of the more realistic iso-energy density hypersurface in the IEF scenario, the results are in agreement with the experimental data.

\begin{figure*}
\includegraphics[scale=0.35]{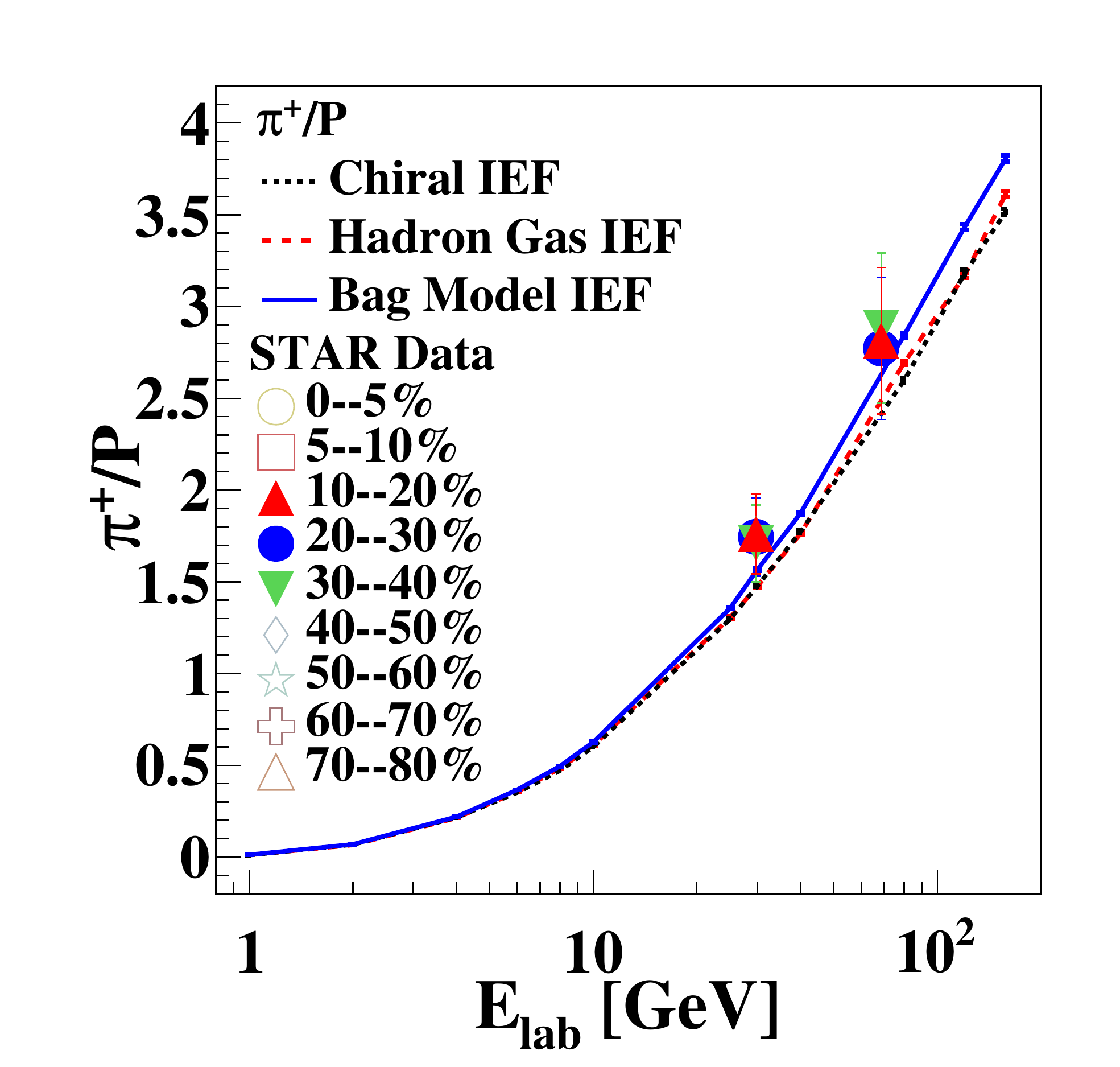}
\includegraphics[scale=0.35]{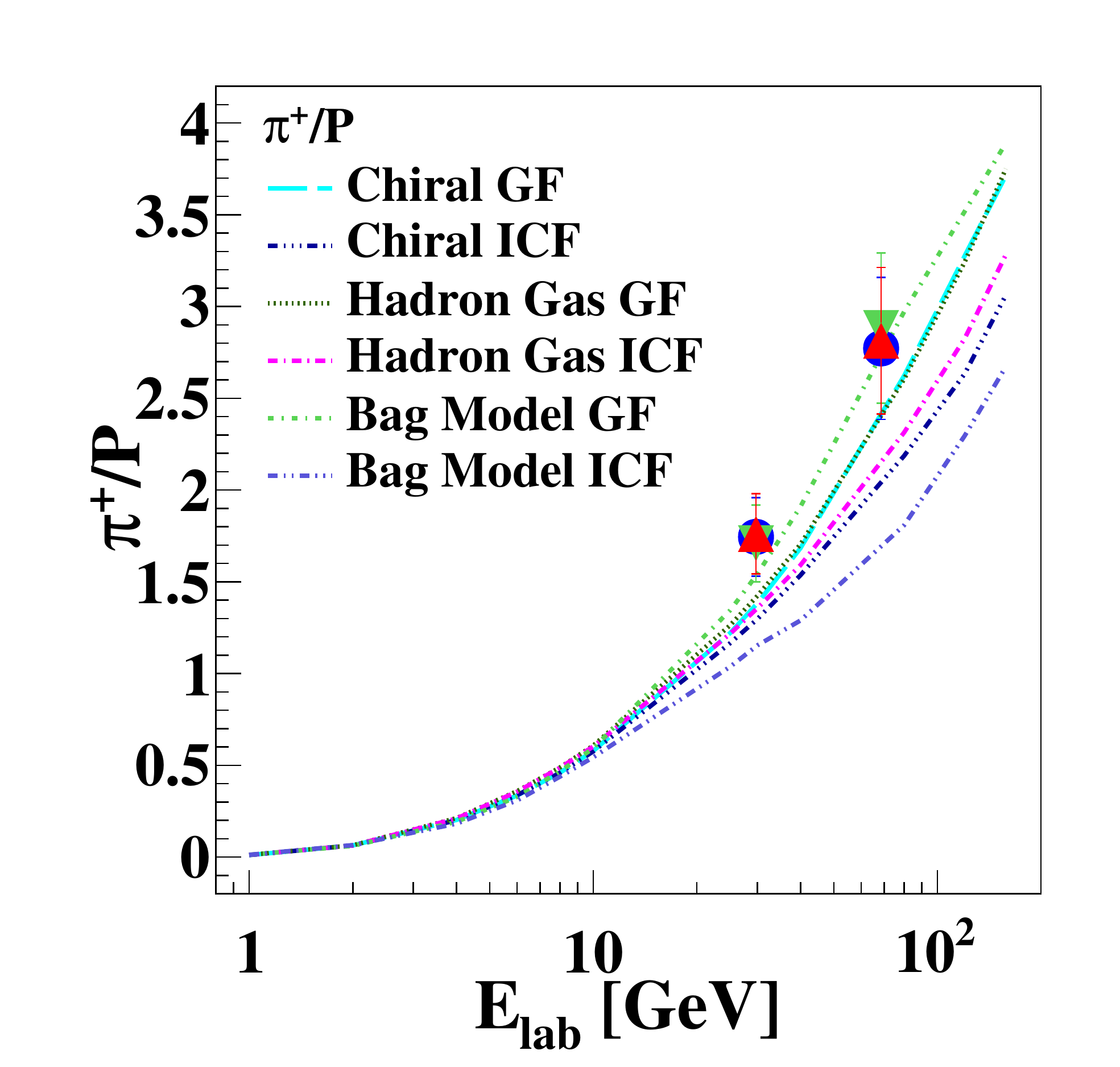}\\
\includegraphics[scale=0.35]{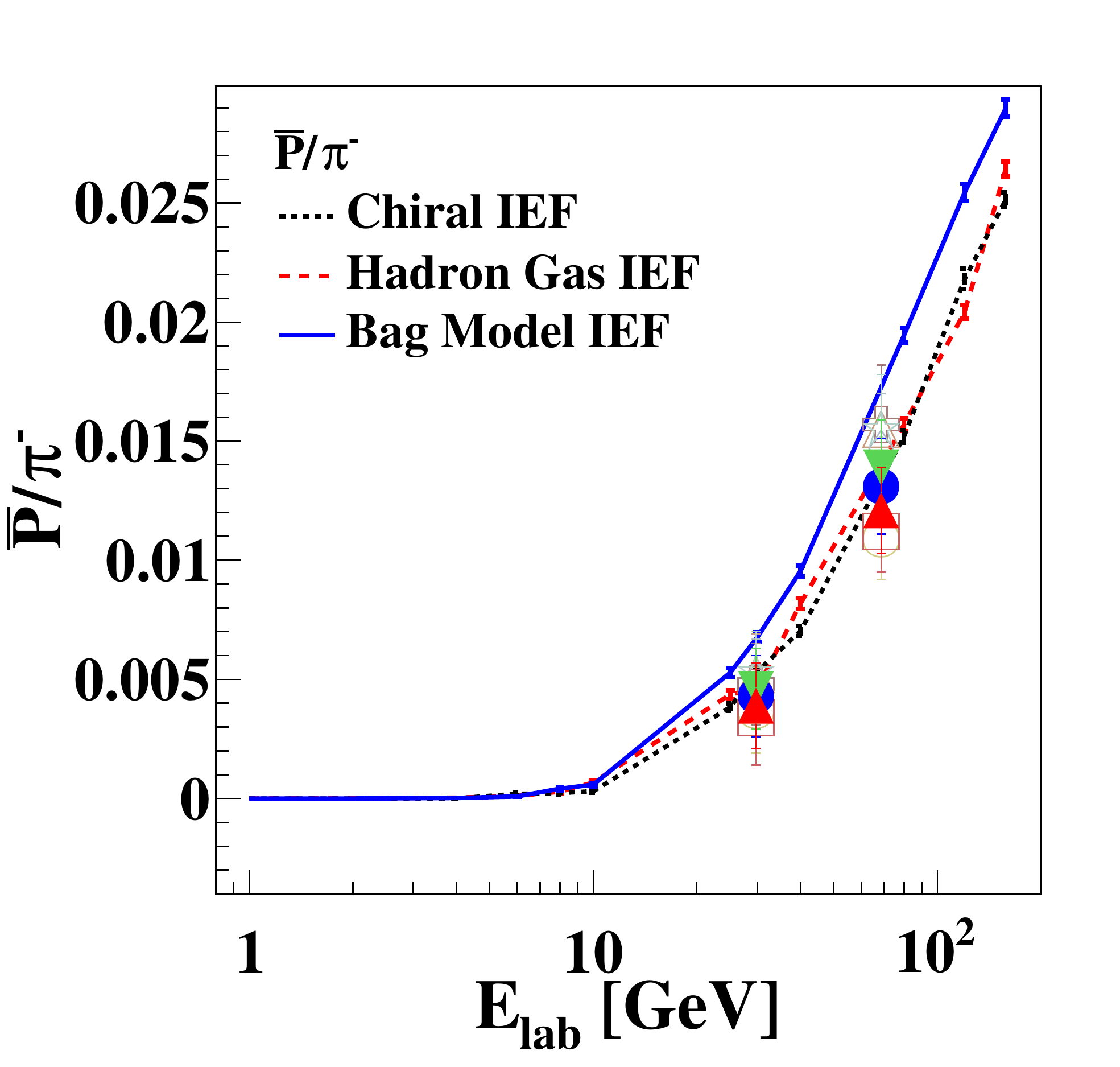}
\includegraphics[scale=0.35]{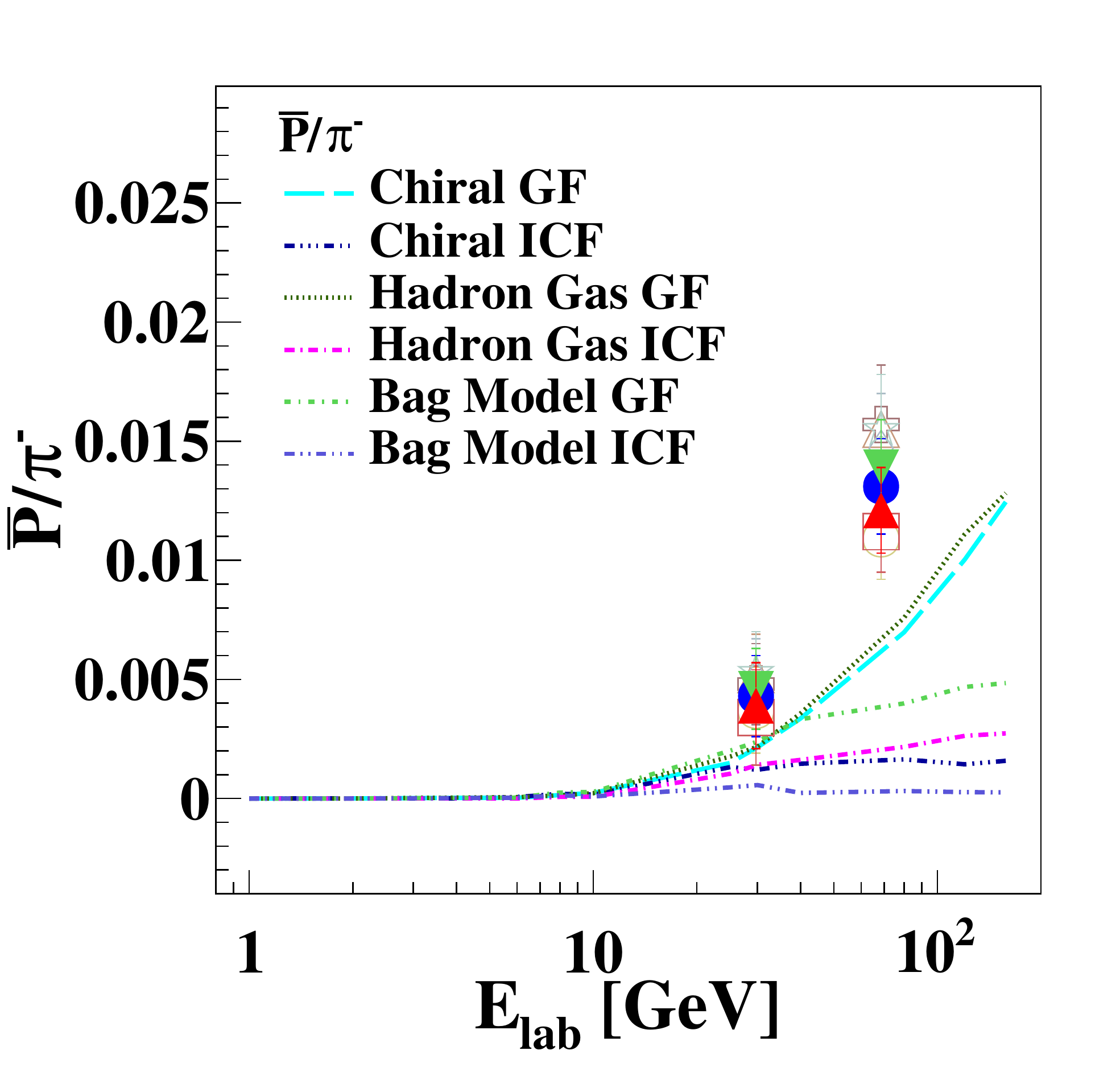}\\
\caption{(Color online) $\pi^{+}$ to proton and anti-proton to $\pi^{-}$ ratio as a function of beam energy for different EoS and particlization modes of UrQMD for non-central (b = 5-9 fm corresponds to approximately 10-40\% central) Au-Au collisions and their comparison with STAR experimental measurements~\cite{Adamczyk:2017iwn} in Au-Au collisions for all available centralities. Vertical bars on the data denote statistical uncertainties.}
\label{fig6}
\end{figure*}

\begin{figure*}
\includegraphics[scale=0.42]{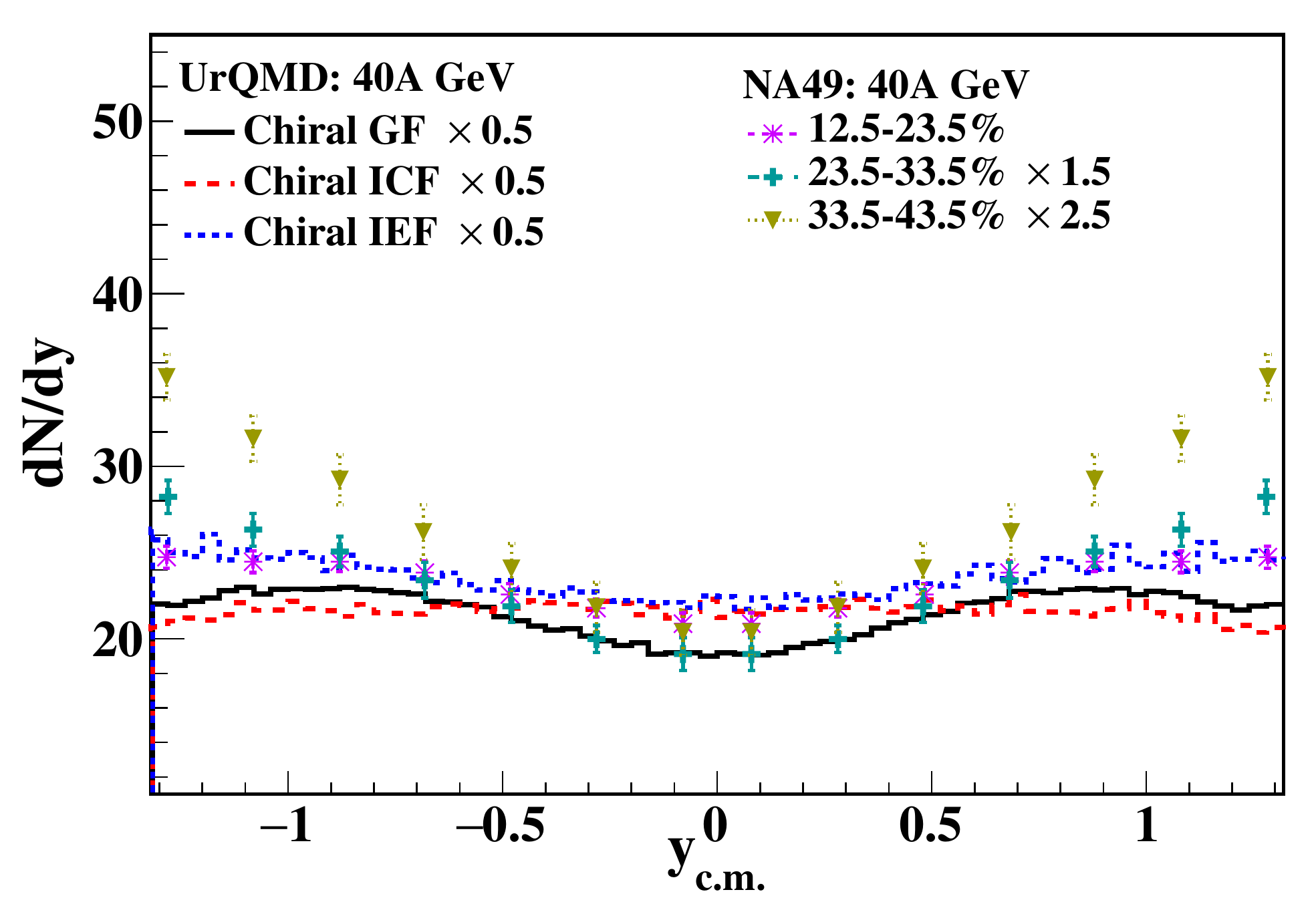}
\includegraphics[scale=0.42]{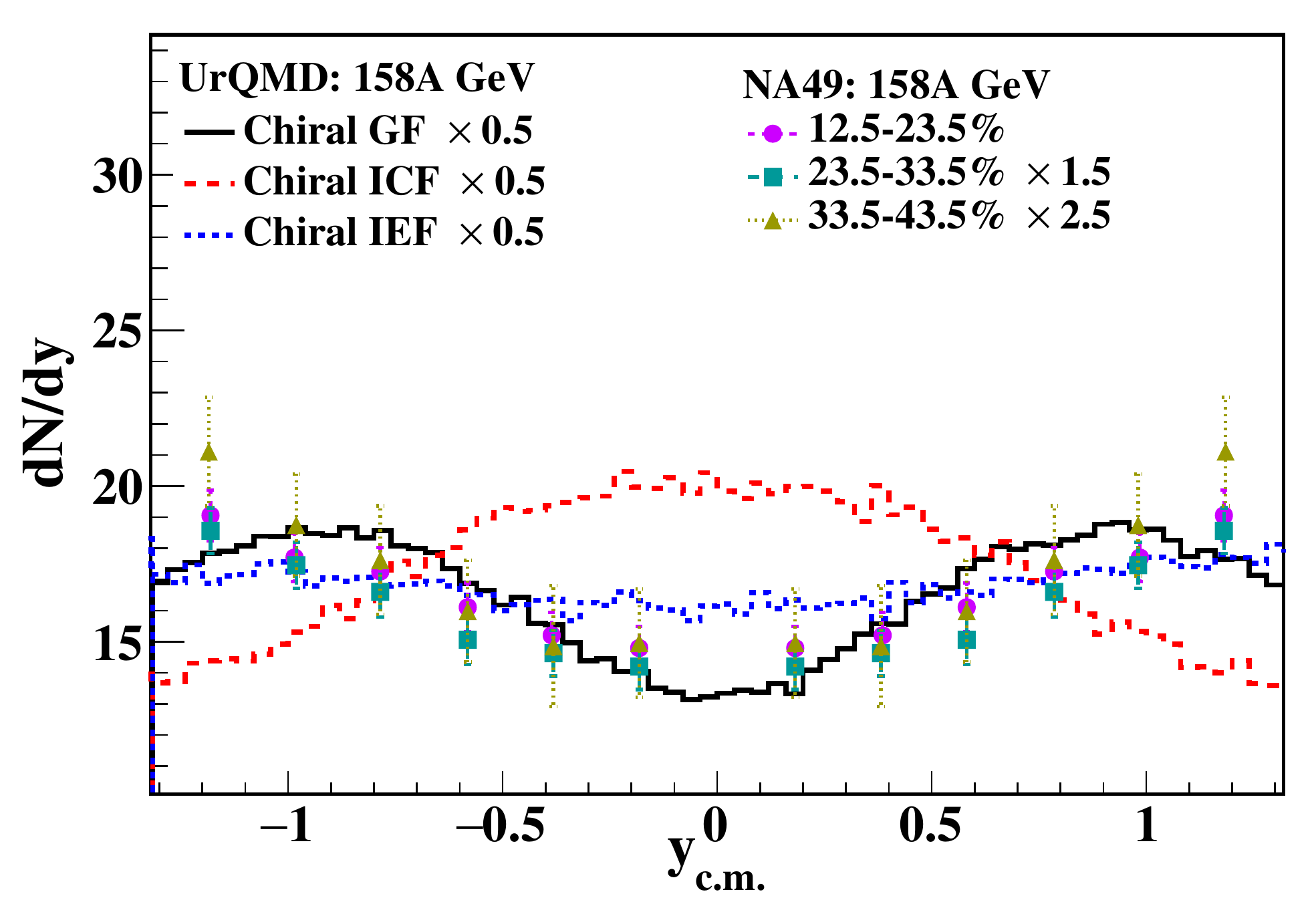}\\
\includegraphics[scale=0.42]{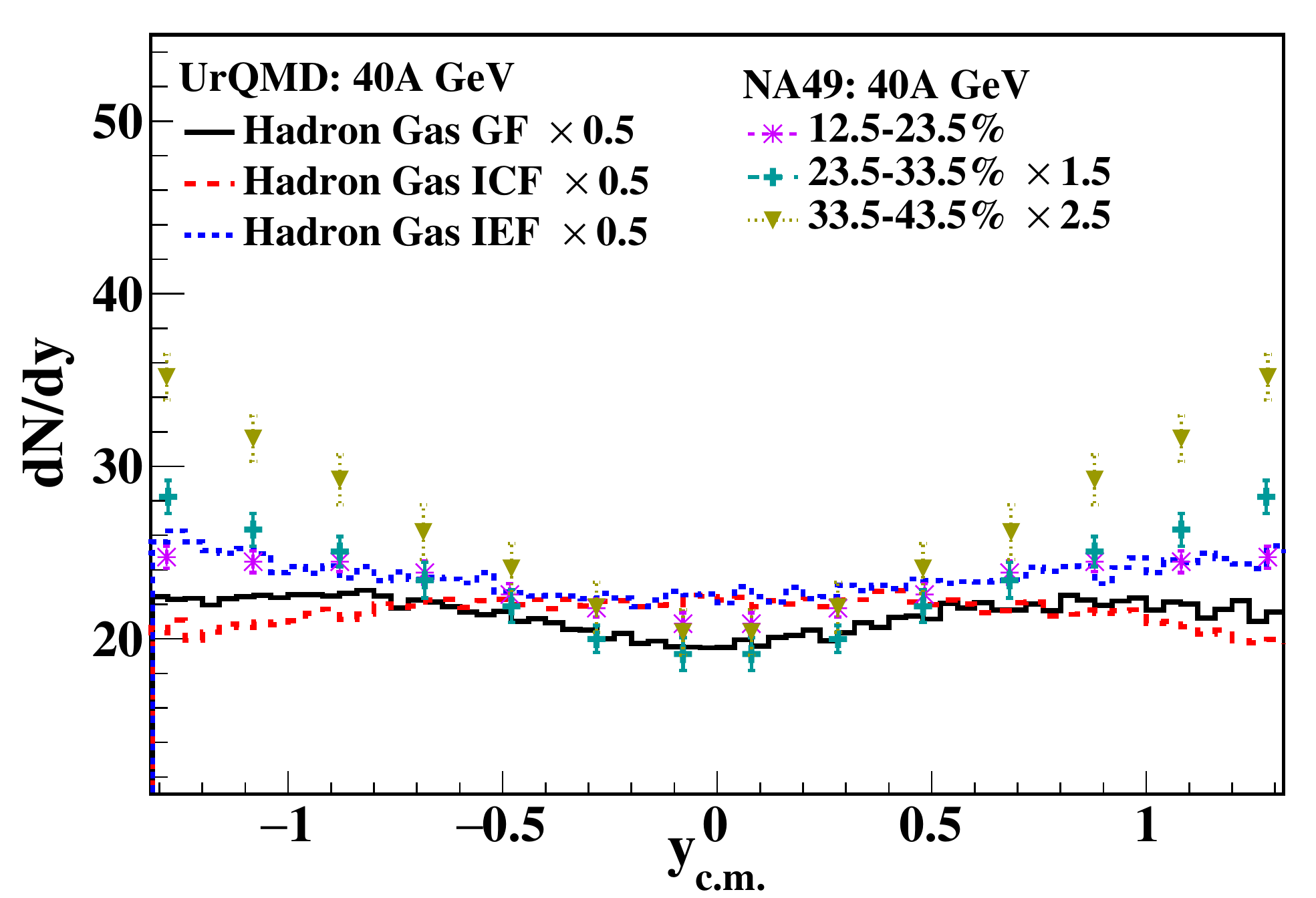}
\includegraphics[scale=0.42]{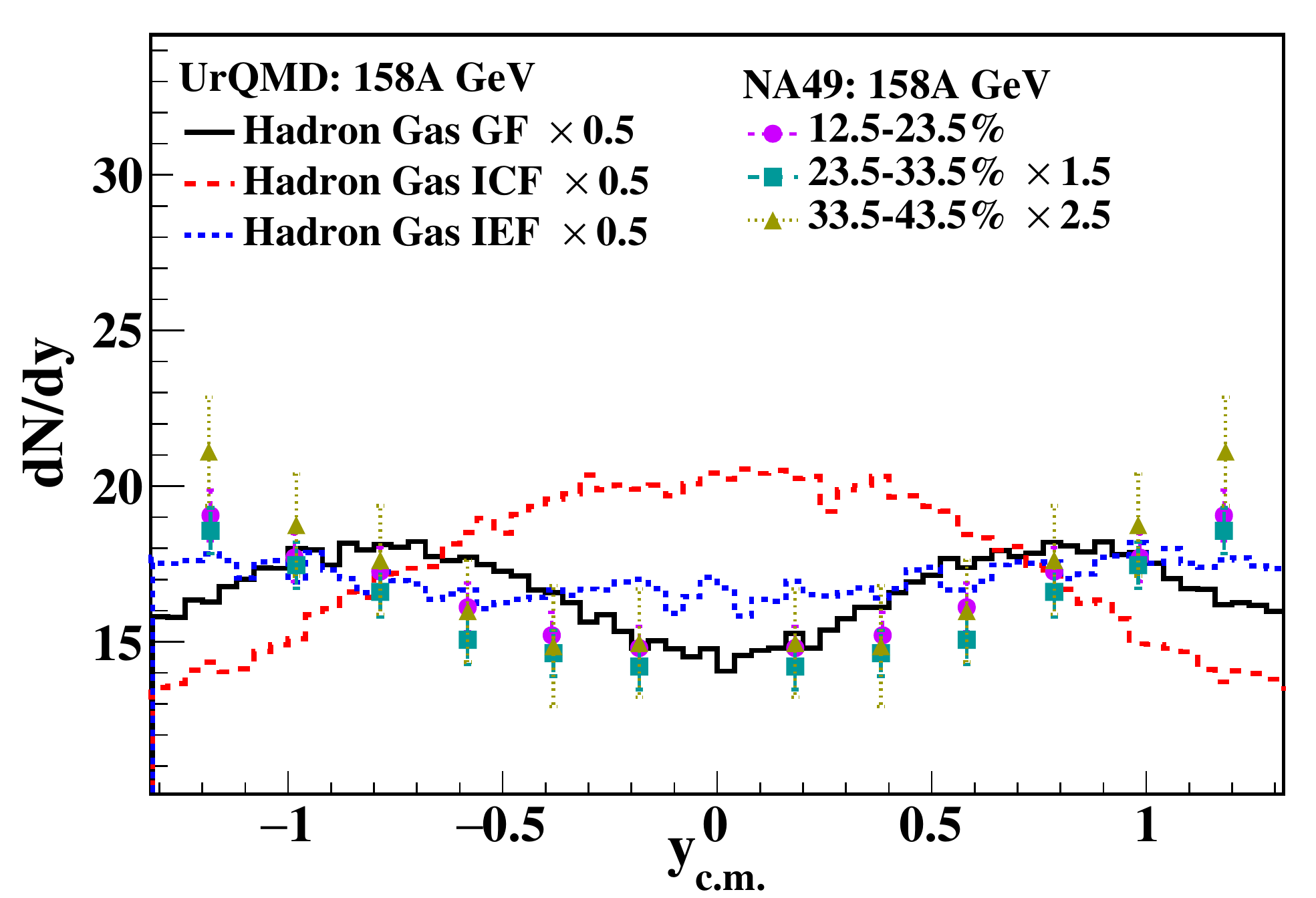}\\
\includegraphics[scale=0.42]{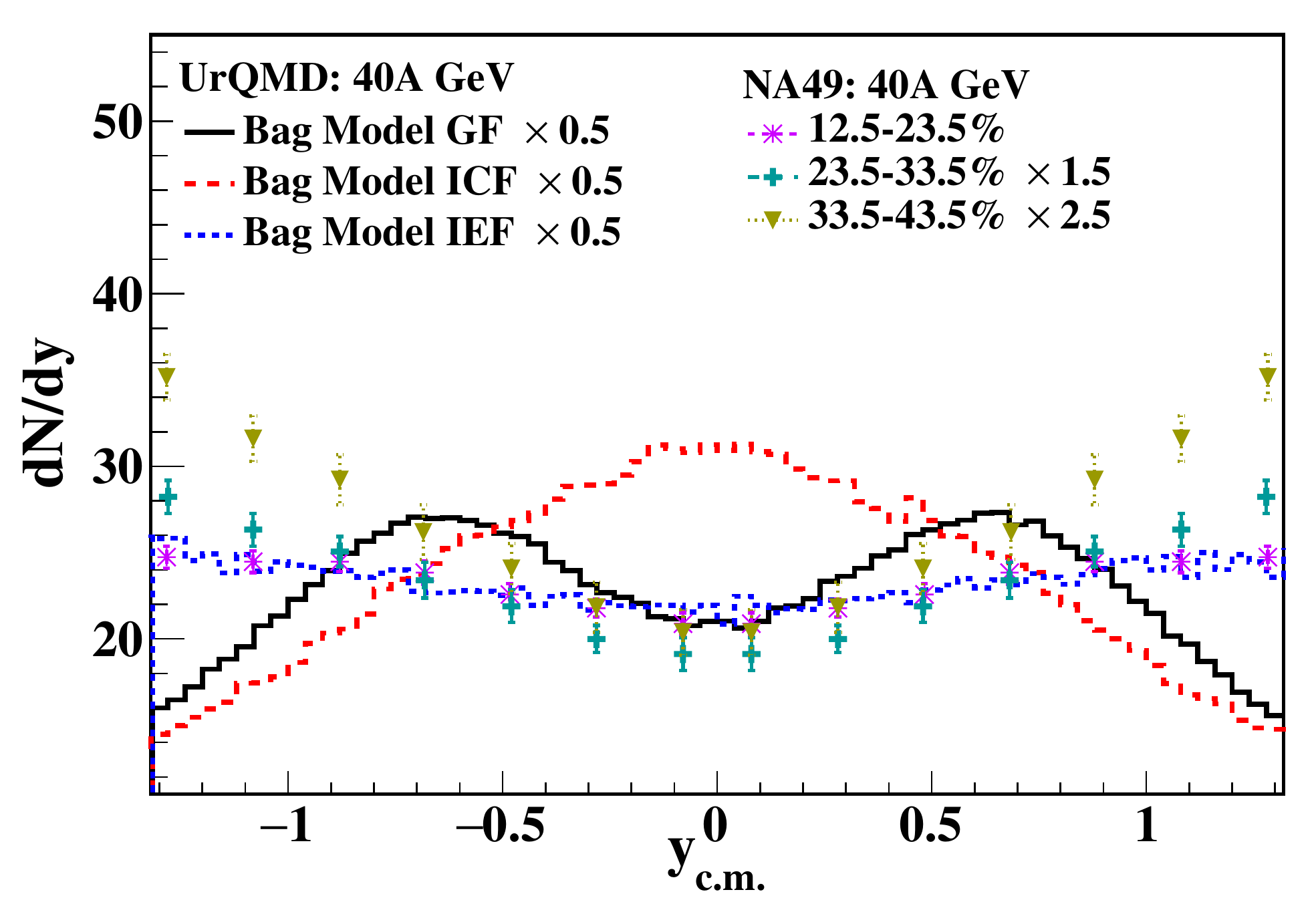}
\includegraphics[scale=0.42]{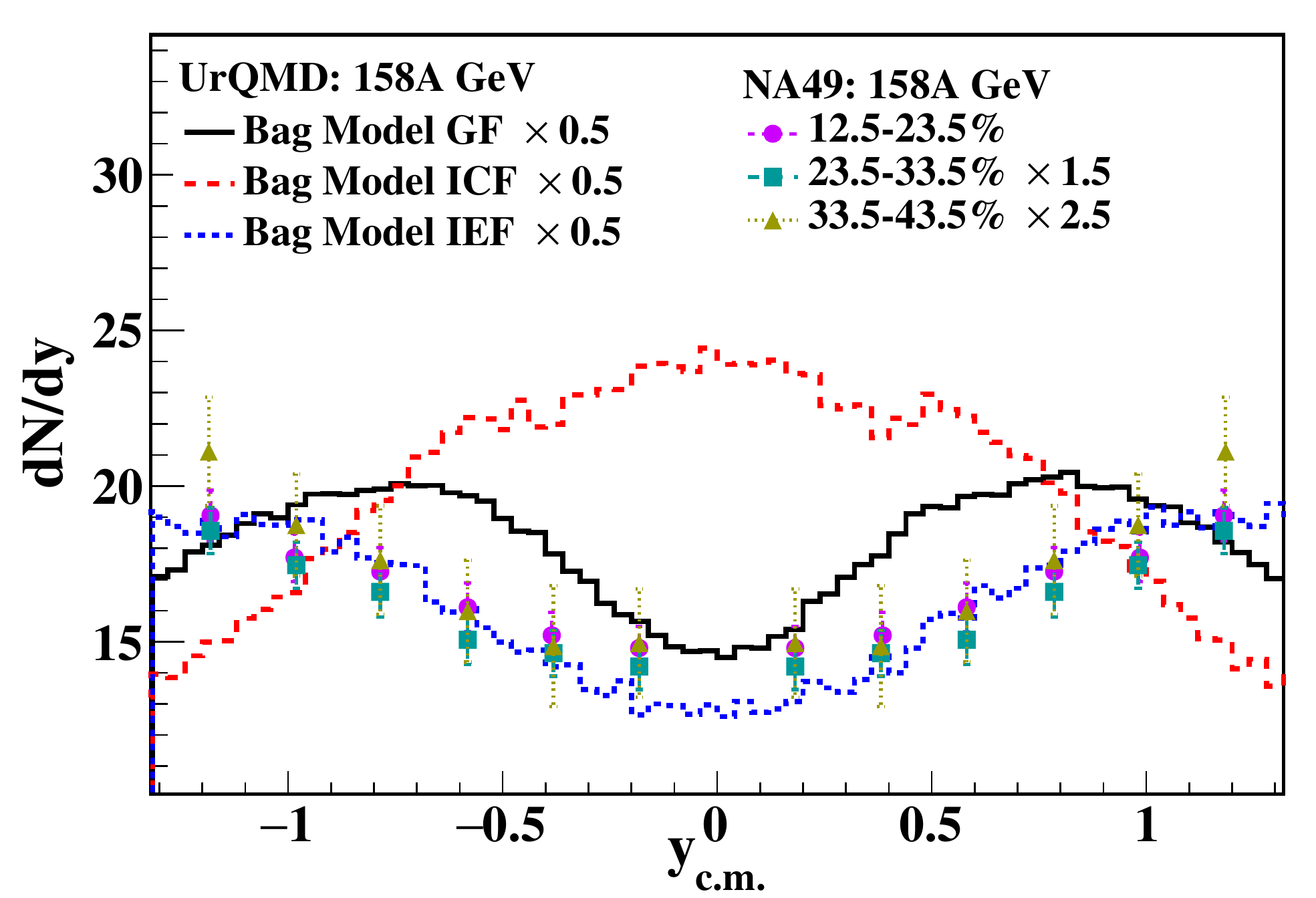}\\
\caption{(Color online) Rapidity spectra of net protons at 40 and 158A GeV beam energies for different EoS and particlization modes of UrQMD for non-central (b = 5-9 fm corresponds to approximately 10-40\% central) Au-Au collisions and their comparison with the measured rapidity spectra of net protons in Au-Au and Pb-Pb collisions by E917~\cite{Back:2000ru} and NA49~\cite{Anticic:2010mp} Collaborations, respectively. Both simulation results and measurements are scaled for better
visualization. Vertical bars on the data denote statistical uncertainties.}
\label{fig6proton}
\end{figure*}

\subsection*{B. Particle ratios}  
In this subsection, we study the effect of different particlization modes under different EoS on particle production in the final state. Particle ratios might show sensitivity to the underlying particlization mode as various criteria for switching from fluid-based to particle-based description can alter the chemical composition of the system. Various particle ratios such as $\rm K^{+}/\pi^{+}$ have been studied in the literature, especially in central heavy-ion collisions, and are believed to be crucial to providing critical information about the medium, such as the onset of deconfinement~\cite{Alt:2007aa}. Investigation of such ratios in noncentral collisions can be exciting and might infer the medium properties.

We start with the strange-to-non-strange particle ratios, such as $\rm K^{-}/\pi^{-}$ and $\rm K^{+}/\pi^{+}$ which is shown in Fig.~\ref{fig4} as a function of beam energies. The ratios are obtained for various particlization scenarios as well as EoS and compared with experimental data. As expected, the ratios are sensitive to the underlying particlization scenario. The unavailability of the data in the desired centrality classes mandates us to compare in three different classes in NA49~\cite{NA49:2002pzu,Alt:2006dk} and STAR experiments~\cite{Adamczyk:2017iwn} which covers the impact parameter region under study. Moreover, the ratio was also compared with measurements in various centrality classes to understand the centrality and beam energy dependence. When ICF scenario is used, there is a non-monotonous trend in the ratio $\rm K^{-}/\pi^{-}$. 
 
 The particle to anti-particle ratios such as $\rm K^{-}/K^{+}$, $\pi^{-}/\pi^{+}$ and $\rm \overline{P}/P$ are estimated as a function of beam energy and shown in Fig.~\ref{fig5}. There is a monotonous trend in all three ratios for all EoS and particlization modes cases. The ratios have shown excellent agreement with experimental measurements in the IEF scenario for all three cases of EoS. Moreover, both in GF as well as ICF modes, the ratios $\rm K^{-}/K^{+}$, $\pi^{-}/\pi^{+}$ have shown agreement with data, but the same can not be said in the case of $\rm \overline{P}/P$. The ratios seem more sensitive to the particlization scenarios because of the possible change in the particle chemistry.
 
 We also estimate the baryon to meson ratio, namely, $\rm P/\pi^{+}$ (shown as the inverse for better visualization) and $\rm \overline{P}/\pi^{-}$ which is shown in Fig.~\ref{fig6}. Agreement with the data is also seen in the case of IEF scenario for all EoS as depicted in Fig.~\ref{fig6}. In the case of $\rm \pi^{+}/P$, the GF scenario is also able to reproduce the data compared to ICF mode. From both Figs.~\ref{fig5} and \ref{fig6}, it is worth noting that there is almost no anti-proton production in the ICF scenario, even at higher beam energies which seems unrealistic. From this investigation,  it seems evident that the IEF scenario brings more clarity to understanding and interpreting the results.
 
 Before closing, we want to bring to the reader's attention that the comparison of protons from the UrQMD model with the experimental data should be cautiously accepted. In the region of low beam energy, $E_{lab}$ $\lessapprox$ 10A GeV, there is a non-negligible contribution of light nuclei production. Primordial nucleons calculated by the model still have a contribution from bound nucleons. This involved all the proton observables in this article. The way out in such a case is to coalesce final state nucleons in UrQMD into light nuclei. The basic principle for coalescence is to find the nucleon clusters with small momentum and position differences. Such study has its importance; however, it is beyond the reach of this investigation. This may not be as significant for anisotropic flow as for particle ratios since the flow coefficients are independent of proton multiplicity. As for particle ratios, this contribution is important as well as for net-rapidity distributions (see next subsection).
 
\begin{figure*}
\includegraphics[scale=0.42]{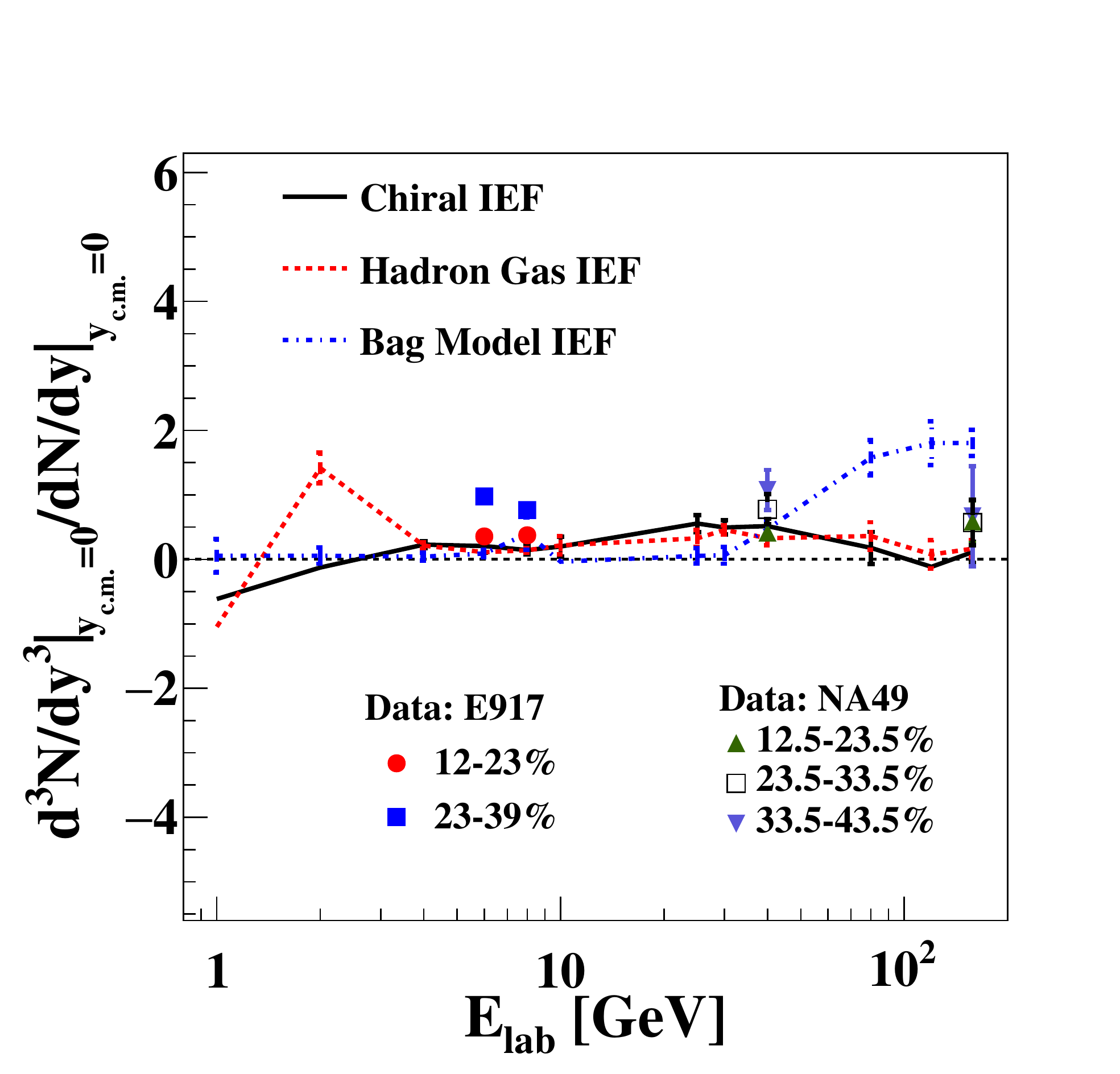}
\includegraphics[scale=0.42]{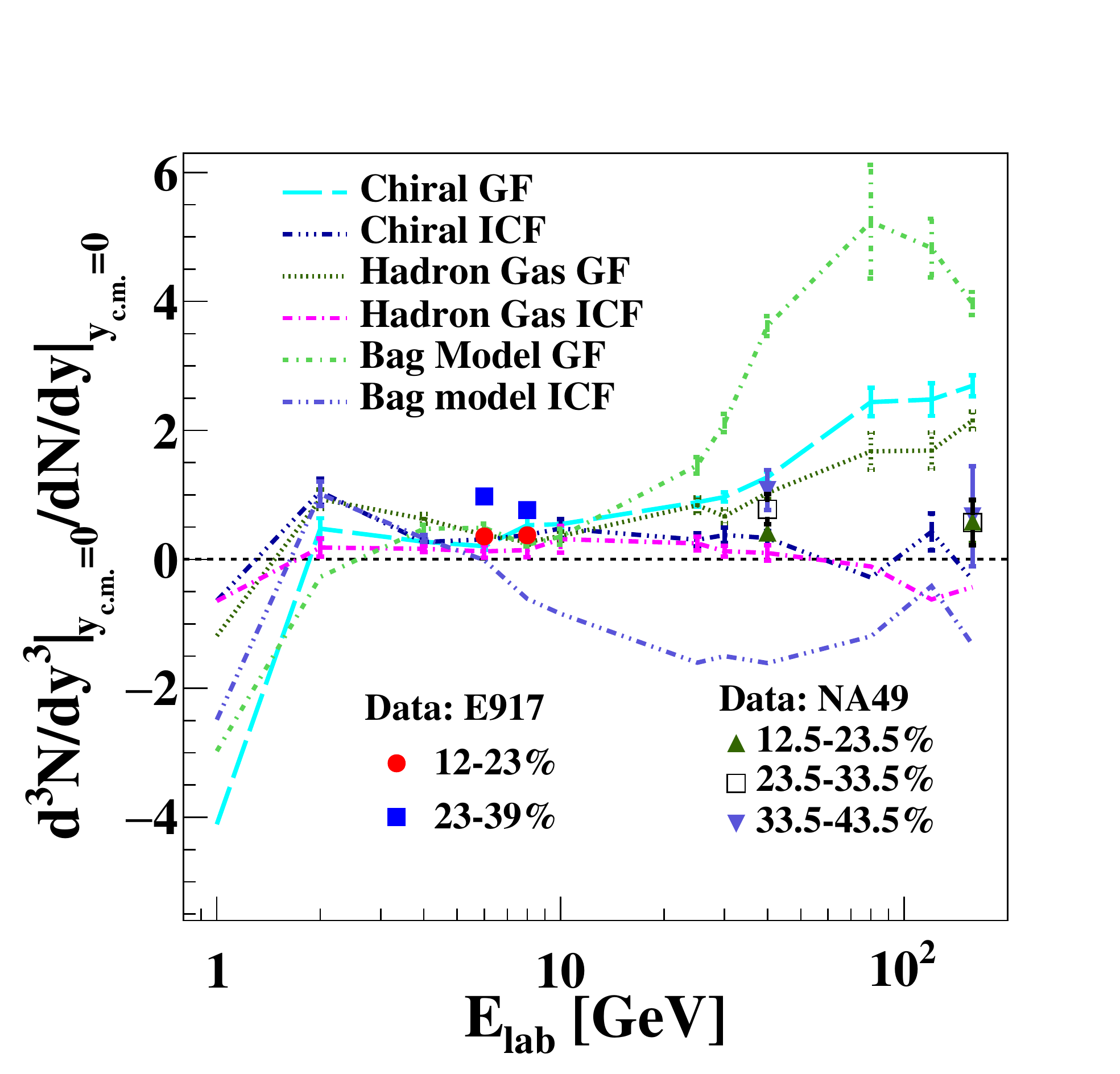}\\
\caption{(Color online) Reduced curvature of rapidity spectra of net protons as a function of beam energy for different EoS and particlization modes of UrQMD for non-central (b = 5-9 fm corresponds to approximately 10-40\% central) Au-Au collisions at midrapidity (-0.5 $<$ $y_{c.m.}$ $<$ 0.5) and its comparison with the calculated reduced curvature of measured rapidity spectra of net protons in noncentral Au-Au and Pb-Pb collisions by E917~\cite{Back:2000ru} and NA49~\cite{Anticic:2010mp} Collaborations, respectively.}
\label{figreduced_curvature}
\end{figure*}

 \subsection*{C. Net-proton rapidity spectra}
 From the rapidity distributions of the net protons, the in-medium properties of stopped protons can be understood. It has been studied as a promising variable for some time, and in the studies performed in Refs.~~\cite{Ivanov:2010cu,Ivanov:2012bh,Ivanov:2013mxa,Ivanov:2015vna,Ivanov:2016xev}, it has been argued that irregularities at the mid-rapidity of the longitudinal spectra of net-protons can be the consequence of the possible onset of the deconfinement. A possible reason behind this may well be the inherited softest point in the nuclear equation-of-state near the phase transition. In the context of this paper, the particlization scenario also can play a crucial role since the production of particles can be sensitive to the fluid to the particle-based description. Since such investigations are primarily performed in the central collision, it is also equally important to study in noncentral collisions. In this investigation, the rapidity distribution of net-proton is estimated for beam energies 1A-158A GeV for various EoS and particlization scenarios. However, the rapidity spectra of net-protons in selected beam energies are shown in Fig.~\ref{fig6proton}. The simulated results are compared to the experimental measurements from E917~\cite{Back:2000ru} and NA49~\cite{Anticic:2010mp} in centralities covering the impact parameter region under investigation. Even at first glance, one can see that rapidity spectra using ICF and GF scenarios are not seem to be in agreement with the measurements. However, the IEF scenario seems to be performing exceptionally well. 
 
 The shape of the rapidity spectra is quantified as a reduced curvature, and its excitation function is studied to gain more insights and have more differential comparisons. The reduced curvature is nothing but global minima or maxima at mid-rapidity, i.e., a double derivative of rapidity spectra at mid-rapidity. This quantity is similar to what was studied in Refs.~\cite{Ivanov:2010cu,Ivanov:2012bh,Ivanov:2013mxa,Ivanov:2015vna,Ivanov:2016xev} and named reduced curvature. In this analysis, the reduced curvature is estimated by fitting the rapidity spectra of net-protons at mid-rapidity using polynomial, which is shown in Fig.~\ref{figreduced_curvature}. The reduced curvature of net-proton rapidity distributions for different particlization models and EoS compared with experimental measurements~\cite{Back:2000ru,Anticic:2010mp}. The reduced curvature for measurements is also estimated in a similar way as for simulations. The results with the IEF scenario for all three cases of EoS show nice agreement with the data. In the ICF scenario, the reduced curvature underestimates the experimental measurements for hadron gas and bag model EoS. In our previous work, we studied it for different EoS with GF scenario, and it disagrees with the data at high beam energies, and the same can also be observed in Fig.~\ref{figreduced_curvature}. One interesting observation from the above figure is that there is very high sensitivity to the particlization scenario in the case of bag model EoS at high beam energies, so much so that we see large negative values of reduced curvature in the case of ICF scenario and large positive values in case GF scenario. We want to mention that the authors of refs~\cite{Ivanov:2010cu,Ivanov:2012bh,Ivanov:2013mxa,Ivanov:2015vna,Ivanov:2016xev} have subtracted the contribution of nucleons bound in nuclei from primordial nucleons, which we have not done in our analysis. Therefore, our results suffer from uncertainties due to the contribution of bound protons.
 
\section*{IV. Summary}

In this article, we have investigated different observables in noncentral Au + Au collisions in beam energy range 1A-158A GeV for various particlization modes and nuclear EoS. In this investigation, we have employed three EoS, namely, hadron gas, bag model, and chiral + deconfinement EoS with particlization models such as gradual (GF), isochronous (ICF), and iso-energy density (IEF) particlization scenario. We started with an anisotropic flow coefficient study for various particlization models for various EoS. We observed that irrespective of any EoS, IEF particlization scenario provides a compelling and qualitative description of the experimental measurements in contrast with the other two scenarios. In particular, experimental measurements of elliptic flow protons and pions were explained well using the IEF scenario at high beam energies. We also observed that directed flow using the IEF scenario does not have any "wiggle" structure in contrast to the other two cases. The slope of the directed flow was examined as a function of beam energies for various particlization modes and EoS, and it was seen that the IEF scenario helped in reproducing the trend of the experimental data qualitatively.

Next, we investigated the effect of particlization models on particle production by estimating the particle ratios as the choice of particlization mode can alter the particle chemistry of the system. We have observed that the strange to non-strange ratio is qualitatively described by IEF and ICF scenarios for all EoS cases. Moreover, in the case of the particle to anti-particle and baryon to meson ratios, the IEF scenario describes the data reasonably well compared to the other two scenarios. Surprisingly, we observed that there was almost no anti-proton production at high beam energies for the ICF scenario, which is unrealistic. Although, more studies in this direction could be helpful.  

In the final section, we investigated the reduced curvature of net protons, which is basically the quantification shape of rapidity at mid-rapidity as a function of beam energy. The reduced curvature obtained for the IEF scenario gave consistent results with experimental measurement, even at high beam energies. The results here are essential and would be helpful once more precise data become available in the upcoming future.

From this investigation, the particlization model-dependent study has given more clarity on the suitability of the EoS in the UrQMD model to examine various experimental observables. We have seen that the choice IEF particlization scenario should be more realistic as it provides compelling agreement with experimental results without being too sensitive to the underlying EoS. However, if one were to choose among the EoS, then both hadron gas and chiral EoS with the IEF scenario would make a good choice for a more realistic combination. Examining other various observables using these combinations would be helpful to make any strong claim.

\section*{Acknowledgements}
S. K. K. acknowledges the financial support provided by the Council of Scientific and Industrial Research (CSIR) (File No.09/1022(0051)/2018-EMR-I), New Delhi. We acknowledge the computing facility provided by the Grid Computing Facility at VECC-Kolkata, India.

\end{document}